\documentclass[12pt]{article}

\usepackage{amsmath,amssymb,amsthm,color,epsfig,mathrsfs}
\usepackage{amsbsy}

\usepackage{tabularx,ragged2e,booktabs,caption}

\newtheorem{theorem}{Theorem}

\newtheorem{proposition}[theorem]{Proposition}

\newcounter{spslist}

\newcounter{geqncount}
    {\refstepcounter{equation}%
     \setcounter{geqncount}{\value{equation}}%
     \setcounter{equation}{0}%
  }%
    {\setcounter{equation}{\value{geqncount}}}

\newcommand{\CC}{\mathbb{C}}
\newcommand{\RR}{\mathbb{R}}
\newcommand{\ZZ}{\mathbb{Z}}
\newcommand{\TT}{\mathbb{T}}
\newcommand{\cchi}[1]{{\textstyle{\chi\big(#1\big)}}}

\newcommand{\Hloc}{H^2_\text{loc}}

\newcommand{\mfh}{\mathfrak{h}}
\newcommand{\mfv}{\mathfrak{v}}
\newcommand{\vv}{{{\mathcal V}_0}}

\renewcommand{\Re}{\mathrm{Re}}
\renewcommand{\Im}{\mathrm{Im}}

\begin{document}

\bibliographystyle{plain} 

\begin{center}
{\bfseries \Large  Spectra of semi-infinite quantum graph tubes}
\end{center}

\vspace{0ex}

\begin{center}
{\scshape \large Stephen P. Shipman \,and\, Jeremy Tillay\\
\vspace{2ex}
{
\itshape
Department of Mathematics\\
Louisiana State University\\
Baton Rouge, Louisiana \ 70803, USA
}}
\end{center}

\vspace{3ex}
\centerline{\parbox{0.9\textwidth}{
{\bf Abstract.}\
The spectrum of a semi-infinite quantum graph tube with square period cells is analyzed.  The structure is obtained by rolling up a doubly periodic quantum graph into a tube along a period vector and then retaining only a semi-infinite half of the tube.  The eigenfunctions associated to the spectrum of the half-tube involve all Floquet modes of the full tube.  This requires solving the complex dispersion relation $D(\lambda,k_1,k_2)=0$ with $(k_1,k_2)\in(\mathbb{C}/2\pi\mathbb{Z})^2$ subject to the constraint $\alpha k_1 + \beta k_2 \equiv 0$ (mod $2\pi$), where $\alpha$ and $\beta$ are integers.  The number of Floquet modes for a given $\lambda\in\mathbb{R}$ \,is\, $2\max\left\{ \alpha, \beta \right\}$.  Rightward and leftward modes are determined according to an indefinite energy flux form.  The spectrum may contain eigenvalues that depend on the boundary conditions, and some eigenvalues may be embedded in the continuous spectrum.
}}

\vspace{3ex}
\noindent
\begin{mbox}
\end{mbox}
\vspace{3ex}

\hrule
\vspace{3ex}


This work investigates the spectrum of a semi-infinite periodic quantum graph tube (half-tube), obtained by rolling a doubly periodic quantum graph into a tube and then cutting it and retaining only one half.  A quantum graph is a metric graph endowed with a Schr\"odinger operator on the edges and supplemented with self-adjoint vertex conditions.

The spectrum of a full (infinite in both directions) quantum graph tube formed from a single-layer hexagonal quantum graph was investigated by Korotyaev and Lobanov for ``zigzag" tubes \cite{KorotyaevLobanov2006,KorotyaevLobanov2007} and by Kuchment and Post in general \cite{KuchmentPost2007}.  In short, the energy $\lambda$ and Bloch wave vector $(k_1,k_2)$ of any eigenfunction (Floquet mode) of the doubly periodic graph are related by a dispersion relation $D(\lambda,k_1,k_2)=0$.  Its spectrum consists of all $\lambda\in\RR$ such that $D(\lambda,k_1,k_2)=0$ for some $(k_1,k_2)$ in the torus $\RR^2/(2\pi\ZZ)^2$, corresponding to a bounded Floquet, or Bloch, mode.  To obtain the spectrum of the tube, the wavevector $(k_1,k_2)$ must be restricted to a rational line $\alpha k_1 + \beta k_2 \equiv 0\;\text{(mod $2\pi$)}$ on the torus, where $\alpha$ and $\beta$ are integers.

Characterizing the spectrum of a semi-infinite tube involves all of the Floquet modes of the tube for each real energy $\lambda$, not only the bounded Floquet modes.
A study of all the the Floquet modes involves considering the relation $D(\lambda,k_1,k_2)=0$ for real $\lambda$ and {\em complex} $(k_1,k_2)$, restricted by the complex relation $\alpha k_1 + \beta k_2 \equiv 0\;\text{(mod $2\pi$)}$.

The reason for involving all Floquet modes is that, when waves are reflected from the boundary where the tube was cut, exponentially decaying modes participate in satisfying the boundary conditions.  In addition, there may exist eigenfunctions that decay exponentially into the infinite side of the tube---these are bound states supported by the boundary at discrete energies.  It will be shown that it is possible to choose self-adjoint boundary conditions that allow eigenvalues embedded within the continuous spectrum.  This is not surprising since a quantum graph tube is a one-dimensional structure in the sense that its symmetry group is a finite-index abelian extension of $\ZZ$.  Eigenfunctions for embedded eigenvalues have been constructed for a variety of one-dimensional differential and difference equations with localized defects \cite{AyaCanoZhevandrov2012,ShipmanRibbeckSmith2010,ShipmanWelters2012}, including quantum graphs \cite[\S 1]{Shipman2014}.

The exponential Floquet modes are often referred to as evanescent modes.  Not only are they necessary for describing the extended states associated with reflection from the boundary of a half-tube, as in this paper, but also for scattering of waves by a localized defect in an infinite tube or transmission of waves across a junction between two semi-infinite tubes.

The spectrum of a semi-infinite combinatorial (as opposed to metric) graph tube has been analyzed by Iantchenko and Korotyaev \cite{IantchenkoKorotyaev2010}, where a magnetic Schr\"odinger operator realizes interactions directly between adjacent vertices instead of acting on edges.  That work considers the zigzag half-tube, which is formed by rolling up a hexagonal graph along one of the fundamental vectors of periodicity.
This corresponds to $\alpha=0$ or $\beta=0$ in the relation above, and the symmetry group of the tube is therefore a direct product $\ZZ\times(\ZZ/p\ZZ)$.

This article treats tubes formed by rolling up a doubly periodic quantum graph along a general vector of periodicity for which $\alpha$ and $\beta$ are both nonzero, so that the symmetry group is not a direct product.  It deals with the simplest doubly periodic quantum graph with square period, which consists of the two-dimensional integer lattice with adjacent vertices connected by edges and natural (Neumann) matching conditions imposed at the vertices.  The structure of the Floquet modes of the tube is ascertained by combining two techniques.  One is  converting the operator on the tube into a discrete system possessing an invariant indefinite flux form (sec.~\ref{subsec:eigenspaces}), which is akin to the reduction to a half-infinite periodic Jacobi operator in \cite{IantchenkoKorotyaev2010}.
The other is a direct analysis of the dispersion relation with respect to the group of symmetries of the tube (sec.~\ref{subsec:dispersion}).

\section{Quantum graph tubes} 

\begin{figure}  
\centerline{\scalebox{0.4}{\includegraphics{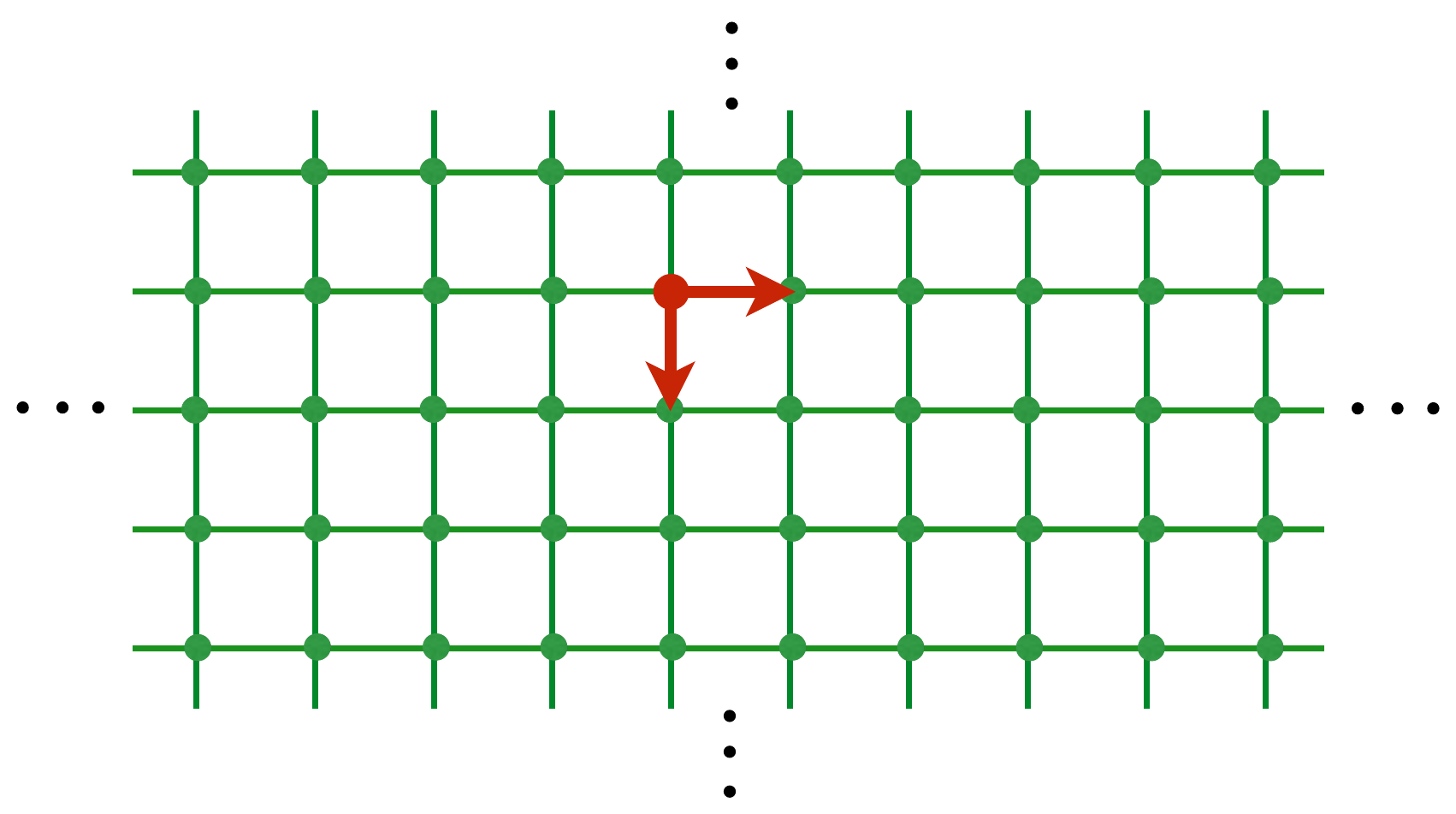}}}
\caption{\small The metric graph $\Gamma$ and a fundamental domain $U$ for the $\ZZ^2$ action.}
\label{fig:gamma}
\end{figure}

Let $\Gamma$ denote the graph shown in Fig.~\ref{fig:gamma}.
The vertex set $\mathcal{V}(\Gamma)$ is the two-dimensional integer lattice, and an edge connects each pair of horizontally or vertically adjacent vertices.  Each edge is identified with the real interval $(0,1)$, which renders $\Gamma$ a metric graph.  The set of edges is denoted by $\mathcal{E}(\Gamma)$, and the set of edges incident with a given vertex $v$ is denoted by $\mathcal{E}_v(\Gamma)$.
This metric graph admits a group of symmetries isomorphic to $\ZZ^2$.  If $\Gamma$ is viewed as a subset of $\RR^2$, an element $(m,n)$ of $\ZZ^2$ acts on $\Gamma$ by 
translating it a distance of $m$ to the right and a distance $n$ upward, and the action can be expressed as $(x,y)\mapsto(x+n,y+m)$ for each $(x,y)\in\Gamma$.
A fundamental domain $U$ consists of one vertex with a horizontal edge and a vertical edge emanating from it.

The metric graph $\Gamma$ becomes a quantum graph when it is endowed with a Schr\"odinger operator $-\partial^2/\partial x^2 + q(x)$ on each edge supplemented with vertex conditions that render it self-adjoint in $L^2(\Gamma)$ (see \cite{BerkolaikoKuchment2013} for details on self-adjoint vertex conditions).  The operator is denoted by $A$:
\begin{align}\label{domA1}
  A \;&=\, -\partial^2/\partial x^2 + q(x) \\
  \mathrm{dom}(A) \;&=\,
  \Bigg\{
     f \in L^2(\Gamma) \cap \bigoplus_{e\in\mathcal{E}(\Gamma)} H^2(e) \,:\,
     \sum_{e\in\mathcal{E}_v(\Gamma)} f_e'(v) =0 \;\;\text{for each $v\in\mathcal{V}(\Gamma)$}
  \Bigg\}.
\end{align}
Here, $x$ is a coordinate on any edge of $\Gamma$, $H^2(e)$ is the Sobolev space of functions on an edge~$e$ whose value and derivative are square integrable.  The condition $\sum_{e\in\mathcal{E}_v} f_e'(v) =0$ is called the Neumann vertex condition, in which $f_e'(v)$ means the derivative of the function $f_e$ at the endpoint $v$, directed into the edge $e$.
In this paper, it is assumed that the potential $q(x_e)$ is the same on each edge, where $x_e$ parameterized an edge $e$ and $q:[0,1]\to\RR$ is a fixed bounded symmetric measurable function,
\begin{eqnarray}
  && q(1-x) = q(x), \\
  && -C<q(x)<C\,,
\end{eqnarray}
for some constant $C$.

Of course, this Schr\"odinger operator acts on the larger space of functions
$f\in\oplus_{e\in\mathcal{E}(\Gamma)} H^2(e)$, still subject to the Neumann vertex conditions, but not required to be in $L^2(\Gamma)$.  On this space, the operator will continue to be denoted by $A$.  The operator $A$ is doubly periodic, that is, it commutes with the $\ZZ^2$ action on functions $f:\Gamma\to\CC$ defined by $f\mapsto f(\cdot+(m,n))$ for $(m,n)\in\ZZ^2$.

\begin{figure}  
\centerline{\scalebox{0.4}{\includegraphics{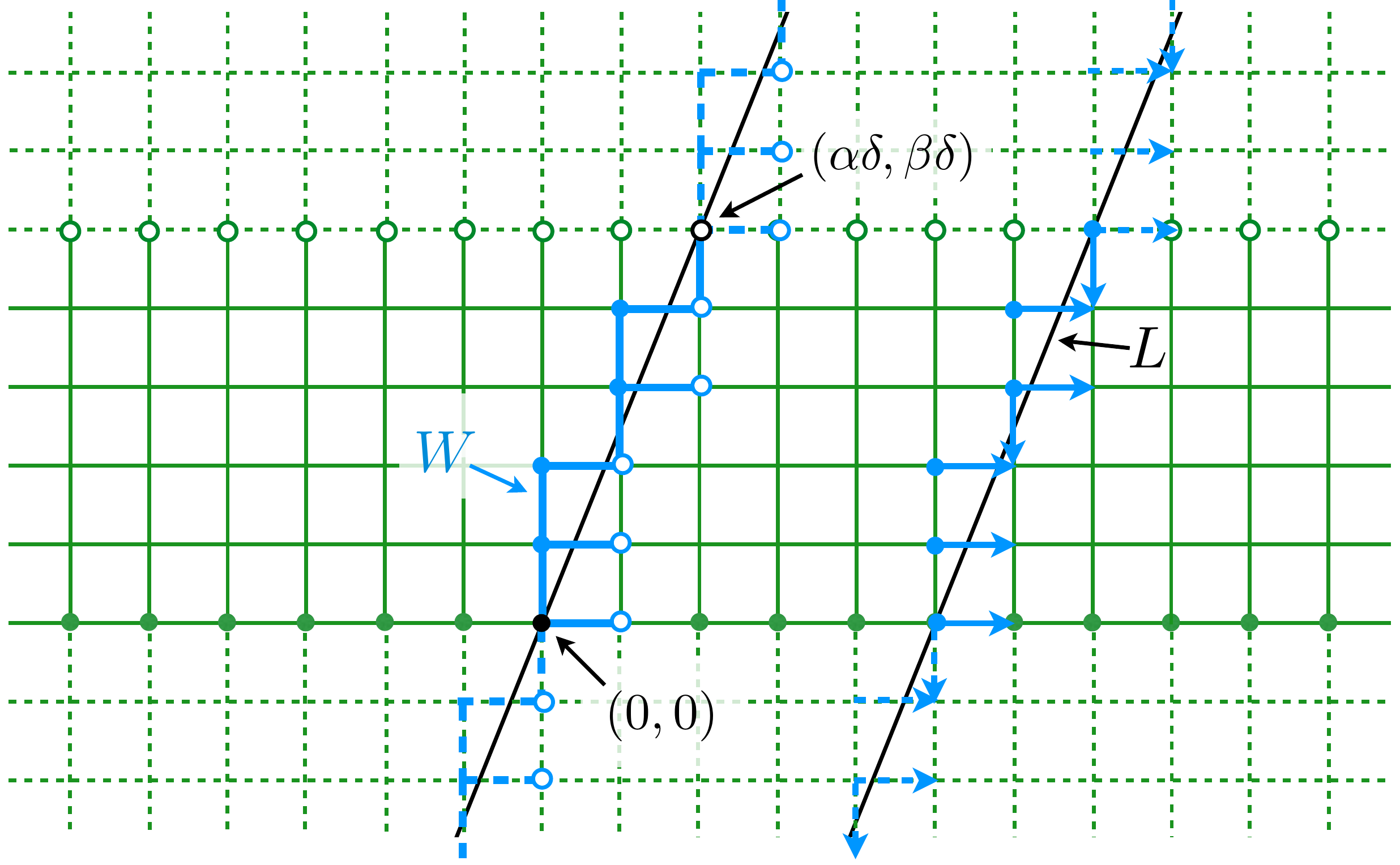}}}
\caption{\small The tube $T$ is formed by identifying points on the square periodic graph $\Gamma$ that differ by the vector $(\alpha\delta,\beta\delta)$.  In this figure, $\alpha=2$, $\beta=5$, and $\delta=1$.  A fundamental domain $W$ (in blue) for the translation action of $\mfh$ on the tube is shown in bold (blue) lines on the left.  The directed edges shown on the right are where the flux is computed to obtain the total flux across the loop $L$ passing around $T$.}
\label{fig:tube}
\end{figure}

A tubular graph is created from $\Gamma$ by considering the action of a subgroup of $\ZZ^2$ and identifying the points belonging to each orbit of this subgroup.  The subgroup is generated by a single element $(\alpha\delta,\beta\delta)$ of $\ZZ^2$; it is denoted by $\langle (\alpha\delta,\beta\delta) \rangle$ and is isomorphic to $\ZZ$.  It will always be assumed that $\alpha$, $\beta$, and $\delta$ are integers such that
\begin{equation}
  \gcd(\alpha,\beta)=1,
  \qquad
  0< \alpha < \beta\,,
  \qquad
  \delta \geq 1\,.
\end{equation}
The tubular graph $T$ formally consists of the orbits of the action of $\langle (\alpha\delta,\beta\delta) \rangle$ on $\Gamma$,
\begin{equation}
  T = \Gamma / \langle (\alpha\delta,\beta\delta) \rangle\,.
\end{equation}
This construction is illustrated in Fig.~\ref{fig:tube}.
The quotient group
\begin{equation}
  G = \ZZ^2 / \langle (\alpha\delta,\beta\delta) \rangle
\end{equation}
acts faithfully as a group of symmetries on $T$ as a metric graph.
$G$ is generated by the cosets of $(1,0)$ and $(0,1)$ of $\langle (\alpha\delta,\beta\delta) \rangle$ in $\ZZ^2$, which are denoted by $\mathfrak{h}$ and $\mathfrak{v}$ (in reference to horizontal and vertical shifts of $\ZZ^2$).
It has the presentation
\begin{equation}
  G = \langle \, \mathfrak{h},\,\mathfrak{v} \,:\,
   \mfh\mfv=\mfv\mfh,\; \mfh^{\alpha\delta} \mfv^{\beta\delta} = \mathfrak{i}\, \rangle,
\end{equation}
in which $\mathfrak{i}$ denotes the identity.
The symmetries $\mathfrak{h}$ and $\mathfrak{v}$ are viewed as a translation together with a rotation of the tube.  Both $\mathfrak{h}$ and $\mathfrak{v}^{-1}$ effect a translation in the same direction along the tube, but they rotate the tube in opposite directions.  The translation-rotation $\mathfrak{h}$ executed $\alpha\delta$ times has the same effect as $\mathfrak{v}^{-1}$ executed $\beta\delta$ times.  $G$ acts on functions by $\mathfrak{g} f (x) = f(\mathfrak{g} x)$ for $\mathfrak{g}\in G$.  

Since $A$ commutes with the $\ZZ^2$ action, it can be considered as an operator on the tube $T$, which can be applied to functions in $\oplus_{e\in\mathcal{E}(T)} H^2(e)$.  In this capacity, it will continue to be denoted by $A$.
When restricted to the domain
\begin{equation}\label{domA2}
  \mathrm{dom}(A) \,=\,
  \Bigg\{
     f \in L^2(T) \cap \bigoplus_{e\in\mathcal{E}(T)} H^2(e) \,:\,
     \sum_{e\in\mathcal{E}_v(T)} f_e'(v) =0 \;\;\text{for each $v\in\mathcal{V}(T)$}
  \Bigg\},
\end{equation}
it is a self-adjoint operator in $L^2(T)$.

\section{Eigenfunctions for the tube}\label{eigenfunctions} 

The operator $A$ on the tube $T$ commutes with the action of~$G$.
Therefore, one seeks functions that are simultaneously eigenfunctions of $A$ and $G$.
These are called the Floquet modes of~$T$.  They are required to satisfy the Neumann vertex conditions but are not required to be in $L^2(T)$, that is, they reside in the space
\begin{equation}
  \Hloc(T) \,=\,
  \Bigg\{
     f \in \bigoplus_{e\in\mathcal{E}(T)} H^2(e) \,:\,
     \sum_{e\in\mathcal{E}_v(T)} f_e'(v) =0 \;\;\text{for each $v\in\mathcal{V}(T)$}
  \Bigg\}.
\end{equation}
Denote the eigenspace of $A$ in $\Hloc(T)$ for eigenvalue $\lambda$ by
\begin{equation}
  V_\lambda \,=\,
  \left\{ f \in \Hloc(T) \,:\, (A-\lambda)\,f = 0 \right\}.
\end{equation}
Since $A$ is self-adjoint, subsequent analysis will be restricted to real values of $\lambda$.

Any function $u\in V_\lambda$ satisfies $-u'' + q(x)u - \lambda u=0$ on each edge, which is identified with the interval $(0,1)$.  Let $c(\lambda,x)$ and $s(\lambda,x)$ be solutions such that
\begin{equation}
\renewcommand{\arraystretch}{1.1}
\left[
\begin{array}{cc}
  c(\lambda,0) & s(\lambda,0) \\
  c'(\lambda,0) & s'(\lambda,0)
\end{array}
\right]
=
\renewcommand{\arraystretch}{1.1}
\left[
\begin{array}{cc}
  1 & 0 \\
  0 & 1
\end{array}
\right],
\end{equation}
in which the the prime denotes the derivative with respect to $x$.  The functions $c(\lambda,x)$ and $s(\lambda,x)$ are necessarily real-valued, and the same pair can be taken for any edge oriented in either direction because of the symmetry of $q(x)$ about $x=1/2$.

The set of energies $\lambda$ for which $s(\lambda,1)=0$ is the {Dirichlet spectrum} of any edge, that is, the Dirichlet spectrum of $-\partial_{xx}+q(x)$ on the interval $(0,1)$,
\begin{equation}\label{sigmaD}
  \sigma_D := \left\{ \lambda\in\RR :  s(\lambda,1)=0 \right\}.
\end{equation}
Each element of $\sigma_D$ is an eigenvalue of infinite multiplicity for $A$ on $\Gamma$ or $T$, and the eigenfunctions are linear combinations of ``loop states" of finite support.  These states will not be analyzed in this paper, as a thorough discussion is available in the literature, particularly in~\cite{Kuchment2005a,KuchmentPost2007} (see Fig.~3 in \cite{KuchmentPost2007}).

Section~\ref{subsec:eigenspaces} develops the structure of $V_\lambda$ by considering the $\ZZ$ action on $T$ generated by $\mfh$ alone.
It is proved that $V_\lambda$ has dimension $2\beta\delta$ and is spanned by generalized eigenfunctions of $\mfh$ in $V_\lambda$.  The corresponding eigenvalues of $\mfh$ are the values of the Floquet multiplier $z_1$ for the eigenvalue $\lambda$ of $A$.
The modes can be classified into $\beta\delta$ rightward modes and $\beta\delta$ leftward modes (Theorem~\ref{thm:modes}).

Floquet modes are also eigenfunctions of $\mfv$, and the corresponding eigenvalue is the Floquet multiplier $z_2$.  Triples $(\lambda,z_1,z_2)$ for which a Floquet mode exists are said to satisfy the dispersion relation for the tube.  This relation is analyzed in section~\ref{subsec:dispersion}.  (More traditionally, the dispersion relation is considered as the corresponding relation between $\lambda$, $k_1$, and $k_2$, where $z_1=e^{ik_1}$ and $z_2=e^{ik_2}$.)

\subsection{Floquet modes and energy flux for the tube}\label{subsec:eigenspaces}

Investigation of the spectrum of the half-tube will proceed from the framework of the problem of reflection (or scattering) of Floquet modes from the boundary of the half-tube, where the infinite tube is cut.
This requires the notion of rightward and leftward modes, which is developed in this section.

The $\lambda$-eigenfunctions $u$ of $A$ will be viewed as solutions of a one-dimensional discrete system of the~form
\begin{equation}\label{dds}
  \mathfrak{h}\,u \,=\, P_\lambda\, u\,,
\end{equation}
in which the operator $P_\lambda$ commutes with the translation $\mathfrak{h}$ and is described below in Proposition~\ref{prop:recurrence}.  The eigenvalues of $P_\lambda$ will then be the eigenvalues of the action of $\mfh$ restricted to $V_\lambda$, and the corresponding eigenfunctions will be Floquet modes of the tube with respect to the symmetry~$\mfh$.  It will be seen in section~\ref{subsec:dispersion} that these Floquet modes are also eigenfunctions of the symmetry $\mfv$.  Much of the structure of the Floquet modes proceeds from the existence of an indefinite energy-flux form that is invariant under $P_\lambda$.

\smallskip
{\em A fundamental domain} $W$ of the action of $\mathfrak{h}$ on $T$ is shown in Fig.~\ref{fig:tube}.
To define it, construct a circle pulled taut around the tube as follows.  Let $L$ be the line in $\RR^2$ passing through $(0,0)$ and $(\alpha\delta,\beta\delta)\in\ZZ^2$.  When taking the quotient with respect to the action of $\langle (\alpha\delta,\,\beta\delta) \rangle$, this line becomes a closed loop (still denoted by $L$) passing around the tube $T$.  Let a ``half-open directed edge" refer to an edge together with a vertex incident to it either at its left endpoint or its upper endpoint, with the edge directed away from its vertex (rightward or downward).  $W$ consists of all half-open edges that intersect $L$, together with all edges that join any two vertices belonging to these half-open edges.
There are $\beta\delta$ vertices contained in $W$, which are denoted by $\left\{ v_k : 0\leq k<\beta\delta \right\}$.

A function $u\in V_\lambda$ is determined by its values on the vertices of $T$ as long as $\lambda\not\in\sigma_D$.  In this case,
%
the Neumann vertex conditions imply a nearest-neighbor relation between the value of $u$ at a given vertex $p$ and the values of $u$ at the four nearest neighbors of $p$,
\begin{equation}\label{laplacian}
  u(\mfh^{-1}p) + u(\mfh\,p) + u(\mfv^{-1} p) + u(\mfv\,p) \,=\, 4 c(\lambda,1)\,u(p)\,.
\end{equation}
This is proved by observing that, if $u$ satisfies $-u''+q(x)u=\lambda u$ on $(0,1)$, then
$u'(0) = (u(1)-c(\lambda,1)u(0))/s(\lambda,1)$, and applying this to each edge emanating from a given vertex together with the Neumann vertex condition.
The left-hand side of (\ref{laplacian}) is ($4$ times) the discrete Laplacian of $u$ restricted to $\mathcal{V}(T)$~\cite{Chung1997}.  Its spectrum is related to that of $A$ through the nonlinear infinite-to-one map $\lambda\mapsto c(\lambda,1)$; this relation is discussed in \cite{Cattaneo1997,Kuchment2005a}.  

With the relation (\ref{laplacian}), one can prove that $u$ is determined by its values on the $2\beta\delta$ vertices of the closure $\overline W$ of the fundamental domain $W$ (or any of its translates $\mathfrak{h}^m \overline W$---again as long as $\lambda\not\in\sigma_D$), as stated in Proposition~\ref{prop:recurrence}.
Denote the set of $2\beta\delta$ vertices in $\overline W$ by $\mathcal{V}_0$, and denote the restriction of $u$ to the vertices in the $m^\text{th}$ $\mfh$-shift of $\overline W$ by $\underline u(m)$,
\begin{equation}
  \underline u(m) = u|_{\mfh^m\mathcal{V}_0} \,.
\end{equation}
Since $\underline u(m)$ is a function from $\mfh^m\mathcal{V}_0$ to $\CC$, it is convenient to identify it with an element of~$\CC^\vv$, which is isomorphic to~$\CC^{2\beta\delta}$.


\begin{proposition}[The eigenspaces $V_\lambda$]\label{prop:recurrence}
If $\lambda\in\RR\setminus\sigma_D$, the dimension of the eigenspace $V_\lambda$ of $A$ is $2\beta\delta$, and the restriction map
\begin{equation}
  V_\lambda \to \CC^\vv \;::\; u \mapsto u|_\vv
\end{equation}
is a linear bijection.
Thus, given a vector $\underline u_0\in\CC^\vv$, there exists a unique function $u\in\Hloc(T)$ that satisfies
\begin{eqnarray}
  (A-\lambda)u &=& 0 \\
  u|_{\vv} &=& \underline u_0
\end{eqnarray}
and there is an invertible operator $P_\lambda:\CC^\vv\to\CC^\vv$ such that
\begin{equation}
  \underline u(m+1) \,=\, P_\lambda\, \underline u(m)\,.
\end{equation}
\end{proposition}

\begin{proof}
Since $\lambda\not\in\sigma_D$, a continuous function $u:T\to\CC$ that satisfies $-u'' + q(x)u - \lambda u=0$ on each edge of $T$ is determined by its restriction to $\mathcal{V}(T)$.
If the edge $e$ of $T$ is directed from vertex $p_0$ to vertex $p_1$, then one computes that $s(\lambda,1) u'(p_0) = u(p_1) - c(\lambda,1) u(0)$.  From this, obtains that for any vertex $p$ of $T$,
\begin{multline}\label{uvertices}
s(\lambda,1) \big( u'(\mfh^{-1}p) + u'(\mfh\,p) + u'(\mfv^{-1} p) + u'(\mfv\,p) \big)
\,=\,  \\
u(\mfh^{-1}p) + u(\mfh\,p) + u(\mfv^{-1} p) + u(\mfv\,p) - 4\,c(\lambda,1)\, u(p).
\end{multline}
This implies that $u$ satisfies the Neumann boundary condition if and only if $u$ restricted to the vertices of $T$ satisfies the nearest-neighbor relation (\ref{laplacian}).  Thus a function $u\in\Hloc(T)$ satisfies $(A-\lambda)u=0$ if and only if $u|_{\mathcal{V}(T)}$ satisfies (\ref{laplacian}).  On the other hand, a function $u:\mathcal{V}(T)\to\CC$ can be extended uniquely to a function $u:T\to\CC$ that satisfies $-u'' + q(x)u - \lambda u=0$ on each edge.  One concludes that the operation of restricting a function's domain from $T$ to $\mathcal{V}(T)$ is a bijective linear map from $V_\lambda$ to the vector space of solutions to the discrete equation (\ref{laplacian}).

Given a function $u:\mathcal{V}(T)\to\CC$ that satisfies (\ref{laplacian}), one can show with the aid of Fig.~\ref{fig:tube}, that the values of $u$ on $\mfh^{m}\mathcal{V}_0$ are determined as a linear function of the values of $u$ on $\mfh^{m-1}\mathcal{V}_0$ and as another linear function of the values of $u$ on $\mfh^{m+1}\mathcal{V}_0$.  Since the relation (\ref{laplacian}) is $\mfh$-invariant, these two operators are inverses of one another.     This, together with the fact that $u$ is the restriction of a unique element $\tilde u$ of $V_\lambda$ to $\mathcal{V}(T)$, establishes the existence of the operator $P_\lambda$ in the Theorem and therewith the uniqueness of the extension $\tilde u\in V_\lambda$.
Existence can be seen by observing that the values of $u$ on $\mfh^{m}\mathcal{V}_0$ can be stipulated arbitrarily.  The details are left to the reader with the aid of Fig.~\ref{fig:tube}.
\end{proof}

{\em A flux form on $T$.}  The structure of the Floquet modes proceeds from an indefinite energy-flux form $[\cdot,\,\cdot]_\lambda$ that is $\mathfrak{h}$-invariant in each of the eigenspaces $V_\lambda$.  It represents the energy flux across a loop running around the tube $T$.
To define it, first consider the cross-flux of two functions $u$ and $v$ along an oriented edge $e$ given~by
\begin{equation}\label{flux1}
  (2i)^{-1} \big( \bar u(x) v'(x) - \bar u'(x) v(x) \big) \,.
\end{equation}
The sign of this flux switches upon reorienting the edge by reparameterizing it in the opposite direction.
If $u$ and $v$ are in $V_\lambda$ for $\lambda\in\RR$, this quantity is constant in $x$; it depends only on the edge $e$ and its orientation.  When $u=v$ it becomes the quadratic form\, $ \Im\big(\bar u(x) u'(x)\big) $.

The loop $L$ intersects $T$ at a finite set $S$ of points in $W$.  Consider the sum of all fluxes along the half-open edges of $W$ that contain a point of $S$, directed rightward or downward, from one side of $L$ to the other.
There are $(\alpha+\beta)\delta$ such edges, as illustrated in Fig.~\ref{fig:tube}.  Since $u$ and $v$ on $W$ are determined by their values on $\vv$, this sum of fluxes is a function of $\underline u(0)$ and $\underline v(0)$.
It will be denoted by $[\underline u(0),\,\underline v(0)]_\lambda$, and its arguments will be considered to lie in $\CC^\vv$.  When this construction is applied to the shifted loop $\mfh^mL$, one obtains a quantity that depends on $\underline u(m)$ and $\underline v(m)$, also denoted by $[\underline u(m),\,\underline v(m)]_\lambda$.


\begin{proposition}[Energy flux form]\label{prop:flux}
For $\lambda\in\RR\setminus\sigma_D$, the flux form $[\cdot,\,\cdot]_\lambda$ in $\CC^\vv$ is an indefinite sesquilinear form that is conjugate-symmetric and has signature $(\beta\delta,\beta\delta)$, that is, it has $\beta\delta$-dimensional positive and negative subspaces.
  
For $u,\,v\in V_\lambda$, the flux $[\underline u(m),\,\underline v(m)]_\lambda$ is invariant in $m$, that is,
\begin{equation}\label{fluxinvariance}
  [\underline u(m),\,\underline v(m)]_\lambda \,=\, [\underline u(0),\,\underline v(0)]_\lambda
\end{equation}
for each integer $m$.
\end{proposition}

\begin{proof}
The cross-flux of two eigenfunctions $u$ and $v$ of $A$ for $\lambda\not=0$ along an edge $e$ directed from vertex $p_0$ to vertex $p_1$ can be written in terms of the values of $u$ and $v$ at $p_0$ and $p_1$,
\begin{equation}\label{flux2}
  \frac{1}{2i} \big( \bar u(x) v'(x) - \bar u'(x) v(x) \big)
  =  \frac{1}{2i\,s(\lambda,1)} \big( \bar u(p_0)\,v(p_1) - \bar u(p_1)\,v(p_0) \big)  \,.
\end{equation}
If the vertices $(m,n)$ in $\vv$ are ordered by $(m_1,n_1)<(m_2,n_2)$ if either $n_1<n_2$ or both $n_1=n_2$ and $m_1<m_2$, and the components of the vectors $\underline u=\underline u(0)$ and $\underline v=\underline v(0)$ are ordered accordingly, one can write the flux $[\underline u,\,\underline v]_\lambda$ as
\begin{equation}
   [\underline u,\,\underline v]_\lambda \,=\, \bar {\underline u}^t J_\lambda\, \underline v\,,
\end{equation}
in which $J$ is a $2\beta\delta\times2\beta\delta$ matrix.
By using the representation (\ref{flux2}) and the diagram of directed edges over which the flux is computed (Fig.~\ref{fig:tube}) to yield $[\underline u,\,\underline v]$, one finds that $J$ is block-diagonal with blocks of the form
$(2\,s(\lambda,1))^{-1}J_{2k}$, where $J_k$ is the $k\times k$ matrix
\begin{equation*}
J_{k} = 
\renewcommand{\arraystretch}{1.1}
\left[
\begin{array}{cccccc}
  0 & -i & & & & \\
  i & 0 & i & & & \\
   & -i & 0 & -i & & \\
   & & -i & \hspace{-3pt}\ddots\vspace{-3pt}\hspace{-3pt} & & \\
   & & & & \hspace{-3pt}\ddots\vspace{-3pt}\hspace{-3pt} & \pm i \\
   & & & & \mp i & 0 \\
\end{array}
\right]\,.
\end{equation*}
The determinants $D_k = \det(J_k - tI)$ satisfy the recursion 
\begin{equation}
  D_k \,=\, t\, D_{k-1} - D_{k-2}\,, \quad
  D_0 = 1\,, \quad
  D_{-1} = 0\,.
\end{equation}
Favard's Theorem guarantees that the solution of this recursion is a sequence of orthogonal polynomials $D_k(t)$ with $k$ distinct real roots.  For even values of $k$, $D_k(t)$ is even, and so the roots come in plus-minus pairs~\cite[Theorems~4.3,\,4.4]{Chihara1978}.  The signature of the quadratic form represented by $J_{2k}$ is therefore $(k,k)$.  One concludes that the signature of $J$ is $(\delta\beta,\delta\beta)$.

The invariance of $[\underline u(m),\,\underline v(m)]$ with respect to $m$ comes from the fact that, for any parallelogram $\mathcal{P}$ in the plane, the sum of the fluxes (\ref{flux1}) along all edges of the graph $\Gamma$ that cross $\mathcal{P}$ is equal to zero.  The edges must be directed into the exterior region of $\mathcal{P}$, and if $\mathcal{P}$ crosses a vertex, one takes the fluxes on the edges incident to that vertex that extend into the exterior region.  The vanishing of the sum of these fluxes is due to the Neumann vertex conditions satisfied by $u$ and $v$ (see~(\ref{domA1})).
The statement of the Theorem follows by applying this fact to the parallelogram with one side connecting $(m_1,0)$ and $(m_1+\alpha\delta,\beta\delta)$ and another side connecting $(m_2,0)$ and $(m_2+\alpha\delta,\beta\delta)$, with $m_1<m_2$.  When folded onto the tube $T$, all flux contributions from the horizontal (top and bottom) sides of $\mathcal{P}$ mutually cancel and the contributions from the two slanted sides are equal to $-[\underline u(m_1),\,\underline v(m_1)]_\lambda$ and $[\underline u(m_2),\,\underline v(m_2)]_\lambda$. 
\end{proof}


\begin{proposition}\label{prop:conservation}
  The propagator operator $P_\lambda$ is flux-unitary, that is,
\begin{equation}
  [P_\lambda\,\underline u\,,\,P_\lambda\,\underline u]_\lambda \,=\, [\underline u,\,\underline u]_\lambda\,.
\end{equation}
\end{proposition}

\begin{proof}
  This follows from Propositions \ref{prop:recurrence} and~\ref{prop:flux}.
\end{proof}

Knowing that $P_\lambda$ is unitary with respect to the indefinite flux form $[\cdot,\,\cdot]_\lambda$, one can ascertain properties of the eigenvalues of $P_\lambda$ and the flux interactions between the corresponding eigenvectors (see the book \cite{GohbergLancasterRodman2005}).  The following theorem deals with the case when $P_\lambda$ is diagonalizable, which is the case except at edges of spectral bands, as will be seen later on.


\begin{theorem}[Structure of eigenmodes]\label{thm:modes}
Let $\lambda\in\RR\setminus\sigma_D$ be given.  Let $w$ be an eigenvalue of $P_\lambda$ with eigenvector $\underline u_0$, and let $u\in V_\lambda$ denote the corresponding eigenmode of $A$.   
The restriction $\underline u(m) = u|_{\mfh^m\vv}$ of $u$ to $\mathcal{V}(T)$,  is given by
\begin{equation}
  \underline u(m) \,=\, w^m\,\underline u_0\,.
\end{equation}
If $P_\lambda$ has $2\beta\delta$ distinct eigenvalues, then $V_\lambda$ is spanned by the corresponding $2\beta\delta$ eigenmodes of~$A$.  The eigenvalues with unit modulus come in conjugate pairs, and the others come in pairs reflected about the unit circle:
\begin{equation}\label{wvalues}
  \renewcommand{\arraystretch}{1.1}
\left.
\begin{array}{lll}
  w_{(j)},\;\bar w_{(j)}\,, & |w_{(j)}| = 1\,, & 0<j\leq\beta'\delta\,, \\
  w_{(j)},\;\bar w_{(j)}^{-1}\,, & |w_{(j)}| < 1\,, & \beta'\delta<j\leq\beta\delta\,. \\
\end{array}
\right.
\end{equation}
Here, $\beta'$ is an integer such that $0\leq\beta'\leq\beta$.
Denote the eigenvectors in $\CC^\vv$ associated with these pairs by $\underline u_{(j1)}$ and $\underline u_{(j2)}$ (with $\underline u_{(j1)}$ corresponding to $w_{(j)}$) and the eigenmodes in $V_\lambda$ by $u_{(j1)}$ and $u_{(j2)}$ ($1\leq j\leq\beta\delta$).  The eigenmodes can be chosen such that the flux interactions between them are
\begin{equation}\label{modefluxes}
  \renewcommand{\arraystretch}{1.1}
\left.
\begin{array}{ll}
  \left[\underline u_{(jk)},\,\underline u_{(j\ell)}\right] = (-1)^\ell\delta_{k\ell}\,, & 0<j\leq\beta'\delta\,, \\
  \left[\underline u_{(jk)},\,\underline u_{(j\ell)}\right] = 1-\delta_{k\ell}\,,& \beta'\delta<j\leq\beta\delta\,, \\
  \left[\underline u_{(j_1k)},\,\underline u_{(j_2\ell)}\right] = 0\,, & j_1\not=j_2\,,
\end{array}
\right.
\end{equation}
with $k,\ell\,=1,2$.
\end{theorem}

\begin{proof}
The first statement comes from Proposition~\ref{prop:recurrence}.

Let $B_\lambda$ be a matrix such that $P_\lambda = \exp(iB_\lambda)$.
Since $P_\lambda$ is unitary with respect to $[\cdot,\,\cdot]$, $B_\lambda$ is self-adjoint with respect to $[\cdot,\,\cdot]$, that is, $[B_\lambda \underline u,\,\underline v]=[\underline u,B_\lambda \underline v]$ for all $\underline u,\underline v\in\CC^\vv$.
Thus, if $\mu$ is an eigenvalue of $B_\lambda$, then so is $\bar\mu$ \cite[Prop.~4.2.3]{GohbergLancasterRodman2005}.
Assuming that the eigenvalues of $P_\lambda$, and therefore also those of $B_\lambda$, are distinct, the structure theorem for operators that are self-adjoint with respect to an indefinite inner product \cite[Theorem 5.1.1]{GohbergLancasterRodman2005} yields a basis $\{\underline u_{(j1)},\,\underline u_{(j2)}\}_{j=1}^{\beta\delta}$ for $\CC^\vv$ consisting of eigenvectors of $B_\lambda$ corresponding to eigenvalues $\{\mu_{(j1)},\,\mu_{(j2)}\}_{j=1}^{\beta\delta}$ with the following property.  For some $\beta':0\leq\beta'\leq\beta$,
\begin{equation}\label{muvalues}
  \renewcommand{\arraystretch}{1.1}
\left.
\begin{array}{ll}
  \mu_{(j1)},\;\mu_{(j2)}\,\in\RR & \text{if}\; 0<j\leq\beta'\delta\,, \\
  \mu_{(j1)} = \bar \mu_{(j2)}\,\not\in\RR & \text{if}\; \beta'\delta<j\leq\beta\delta\,, \\
\end{array}
\right.
\end{equation}
and (\ref{modefluxes}) is satisfied.  Since $P_\lambda = \exp(iB_\lambda)$, the vectors $\underline u_{(j\ell)}$ are eigenvectors of $P_\lambda$ corresponding to eigenvalues $w_{(j\ell)} = \exp(i\mu_{(j\ell)})$ for $1\leq j\leq\beta\delta$ and $\ell=1,2$.  Define $w_{(j)}=w_{(j1)}$ for $1\leq j\leq\beta\delta$.  By the second line of (\ref{muvalues}), one has $\bar w_{(j)}^{-1}=w_{(j2)}$ and $|w_{(j)}|\not=1$ for $\beta'\delta<j\leq\beta\delta$, and this completes the second line of (\ref{wvalues}).  By the first line of (\ref{muvalues}), one has $|w_{(j)}|=1$ for $0<j\leq\beta'\delta$.  Proposition~\ref{prop:zequation} below will establish that the eigenvalues of $P_\lambda$ come in conjugate pairs, and Proposition~\ref{prop:fluxes} will establish that $[\underline u,\,\underline u]$ and $[\underline v,\,\underline v]$ have opposite sign whenever $\underline u$ and $\underline v$ are eigenvectors for complex conjugate modulus-$1$ eigenvalues.  This will allow an arrangement of the eigenvalues for $0<j\leq\beta'\delta$ such that $w_{(j2)}=\bar w_{(j1)}$ and hence $w_{(j2)}=\bar w_{(j)}$, thus completing the first line of (\ref{wvalues}).
\end{proof}

{\em Rightward and leftward Floquet modes.}  Equation (\ref{wvalues}) shows that, generically, the eigenvalues come in conjugate pairs on the unit circle and reflection pairs about the unit circle.  The first type correspond to Floquet modes $u$ that have bounded amplitude, and according to (\ref{modefluxes}), one has $[\underline u,\,\underline u]=1$ and the other has $[\underline u,\,\underline u]=-1$.  These are {\em propagating} modes, or extended states, with one carrying energy to the right and the other carrying energy to the left.  The interaction energy between two different propagating modes is zero.  The second type correspond are pairs of exponential modes; one grows and the other decays as $m$ increases.  These modes carry no energy by themselves but a field that is a superposition of both modes of one pair does carry energy, as the middle equation of (\ref{modefluxes}) shows.  The modes that decay as $m$ increases are rightward evanescent, and those that grow are leftward evanescent.  In summary, a Floquet mode $u$ corresponding to an eigenvalue $w$ of $P_\lambda$ is classified as follows.
\begin{equation}
  \renewcommand{\arraystretch}{1.2}
\left.
  \begin{array}{lll}
    \text{rightward propagating:} & |w|=1, & [\underline u,\,\underline u]>0, \\
    \text{leftward propagating:}   & |w|=1, & [\underline u,\,\underline u]<0, \\
    \text{rightward evanescent:}   & |w|<1, & \\
    \text{leftward evanescent:}   & |w|>1. & \\
  \end{array}
\right.
\end{equation}
For each $j:0<j\leq\beta'\delta$, there is a pair of propagating modes, one rightward and one leftward, corresponding to $w_{(j)}$ and $\bar w_{(j)}$, and for each $j:\beta'\delta<j\leq\beta\delta$, there is a pair of evanescent modes, one rightward and one leftward, corresponding to $w_{(j)}$ and $\bar w_{(j)}^{-1}$.  These modes will be denoted by $u_{(j+)}$ for the rightward one and $u_{(j-)}$ for the leftward one.

\subsection{Complex dispersion relation for the tube}\label{subsec:dispersion}

A direct analysis of the complex dispersion relation reveals additional information about the Floquet modes.
Specifically, it completes the proof of Theorem~\ref{thm:modes} by showing that the unit-modulus Floquet multipliers $w$ come in complex conjugate pairs and that the corresponding eigenvectors $\underline u$ and $\underline v$ are such that  $[\underline u,\,\underline u]_\lambda$ and $[\underline v,\,\underline v]_\lambda$ have opposite sign.
It also shows that any Floquet multiplier of multiplicity greater than one must have modulus $1$ and occurs at a band edge (a local extremum of the graphs in Fig.~\ref{fig:spectrum}).

Each of the Floquet modes described in the previous section is a simultaneous eigenfunction of $A$ and $\mfh$, and together the modes span the eigenspace $V_\lambda$ for $A$.  Because $\mfv$ commutes with both $A$ and $\mfh$, each of these modes $u$ is also an eigenfunction of $\mfv$:
\vspace{-1ex}
\begin{eqnarray}
  A\,u &=& \lambda u\,, \\
  \mathfrak{h}\, u &=& z_1 u\,,\\
  \mathfrak{v}\, u &=& z_2 u\,.
\end{eqnarray}
The Floquet multiplier $z_1$ takes on the values of the eigenvalues $w_{(jk)}$ described in Theorem~\ref{thm:modes}.  The relation between $\lambda$, $z_1$, and $z_2$ is analyzed in this section.

The relation $\mathfrak{h}^{\alpha\delta} \mathfrak{v}^{\beta\delta} = \mathfrak{i}$ implies that $z_1^{\alpha\delta}z_2^{\beta\delta}=1$.
The nearest-neighbor relation (\ref{laplacian}) applied to such $u$ yields the dispersion relation $z_1 + z_1^{-1} + z_2 + z_2^{-1} = 4\,c(\lambda,1)$ for the doubly periodic graph $\Gamma$.  This relation together with $z_1^{\alpha\delta}z_2^{\beta\delta}=1$ comprise the dispersion relation for the tube,
\begin{equation}\label{zequation1}
  \renewcommand{\arraystretch}{1.1}
\left\{
  \begin{array}{l}
    z_1 + z_1^{-1} + z_2 + z_2^{-1} = 4\,c(\lambda,1) \\
    z_1^{\alpha\delta} z_2^{\beta\delta} = 1.
  \end{array}
\right.
\qquad
\text{(dispersion relation for $T$)}
\end{equation}
Notice that, if $(z_1,z_2)$ satisfies this pair, then so does $(z_1^{-1},z_2^{-1})$.

The condition $z_1^{\alpha\delta} z_2^{\beta\delta} = 1$ is equivalent to 
\begin{equation}
  z_1^\alpha z_2^\beta = \cchi{\frac{\ell}{\delta}}
  \quad \text{for some integer } \ell:\,0\leq\ell<\delta,
\end{equation}
in which
\begin{equation}
  \chi(t) := e^{2\pi it}.  
\end{equation}
Thus (\ref{zequation1}) is equivalent to the existence of $\ell:\,0\leq\ell<\delta$ such that
\begin{equation}\label{zequation2}
  \renewcommand{\arraystretch}{1.1}
\left\{
  \begin{array}{l}
    z_1 + z_1^{-1} + z_2 + z_2^{-1} = 4\,c(\lambda,1) \\
    z_1^\alpha z_2^\beta = \chi\!\left( \frac{\ell}{\delta} \right). \\
  \end{array}
\right.
\end{equation}
The following proposition characterizes the solutions of (\ref{zequation2}).

\begin{proposition}\label{prop:zequation}
(a) If $\alpha$ and $\beta$ are positive integers with $\gcd(\alpha,\beta)=1$ and $(z_1,z_2)\in(\CC^*)^2$, then (\ref{zequation2}) holds if and only if there exists $z\in\CC^*$ such that
\begin{equation}\label{zequation3}
  \renewcommand{\arraystretch}{1.1}
\left\{
  \begin{array}{l}
    z^\beta + z^{-\beta} + \eta z^\alpha + \eta^{-1} z^{-\alpha} = 4\,c(\lambda,1),\; \text{ with } \eta=\cchi{\frac{-\ell}{\beta\delta}} \\
    z_1 = z^{\beta} \\
    z_2 = \eta^{-1} z^{-\alpha}\,.
  \end{array}
\right.
\end{equation}
Such $z$ is unique.  The same pair $(z_1,z_2)$ satisfies the modification of the system (\ref{zequation3}) by the replacements
\begin{eqnarray}
  z \,\mapsto\, \cchi{\frac{j}{\beta}}z\,,
  \quad
  \eta \,\mapsto\, \cchi{\frac{-j\alpha}{\beta}}\eta\,,
\end{eqnarray}
featuring the isomorphism $\cchi{\frac{j}{\beta}}\mapsto\cchi{\frac{-j\alpha}{\beta}}$ of the group of $\beta^\text{th}$ roots of $1$.

(b) The replacement
\begin{eqnarray}
  && z \,\mapsto\, z^{-1}\cchi{\frac{j}{\beta}}, \quad j\alpha \equiv 1\, (\mathrm{mod}\,\beta) \\
  && \eta \,\mapsto\, \cchi{\!-\!\frac{\delta-\ell}{\delta\beta}}
\end{eqnarray}
results in the replacement of the Floquet pair
\begin{equation}
  (z_1,z_2) \,\mapsto\, (z_1^{-1},z_2^{-1})\,,
\end{equation}
and the replacement
\begin{eqnarray}
  && z \,\mapsto\, \bar z\, \cchi{\frac{j}{\beta}}, \quad j\alpha \equiv 1\, (\mathrm{mod}\,\beta) \\
  && \eta \,\mapsto\, \cchi{\!-\!\frac{\delta-\ell}{\delta\beta}}
\end{eqnarray}
results in the replacement of the Floquet pair
\begin{equation}
  (z_1,z_2) \,\mapsto\, (\bar z_1,\bar z_2)\,.
\end{equation}

(c) If $z$ satisfies the first equation in the system (\ref{zequation3}) and $|z|\not=1$, then $z$ is a simple solution of that equation.
\end{proposition}

\begin{proof}
To prove that (\ref{zequation3}) implies (\ref{zequation2}) is straightforward.
  To prove the uniqueness of the number $z\in\CC^*$ that satisfies (\ref{zequation3}), suppose $\eta^{-1}z^{-\alpha}=\eta^{-1}w^{-\alpha}$ and $z^\beta=w^\beta$ for some $z$ and $w$ in $\CC^*$.  Then $(zw^{-1})^\alpha=1=(zw^{-1})^\beta$.  Since $\gcd(\alpha,\beta)=1$, $zw^{-1}=1$, so that $z=w$.
  
  To prove the existence of such $z$ under the assumption of (\ref{zequation2}), suppose that $z_1^\alpha z_2^\beta=\zeta:=\cchi{\frac{\ell}{\delta}}$, and let $\eta$ be such that $\eta^{-\beta}=\zeta$.  Then choosing numbers $w_1$ and $w_2$ such that $z_1=w_1^\beta$ and $z_2=\eta^{-1}w_2^{-\alpha}$ yields $(w_1w_2^{-1})^{\alpha\beta}=1$ and therefore $w_1w_2^{-1}=\cchi{\frac{r}{\alpha\beta}}$ for some integer $r$.  Since $\gcd(\alpha,\beta)=1$, there exist integers $m$ and $n$ such that $r=-m\alpha + n\beta$, or
\begin{equation}
  \frac{r}{\alpha\beta} = -\frac{m}{\beta} + \frac{n}{\alpha}\,.
\end{equation}
This implies that
\begin{equation}
  w_1w_2^{-1} = \cchi{\frac{-m}{\beta}} \cchi{\frac{n}{\alpha}},
\end{equation}
so that $w_1\cchi{\frac{m}{\beta}}=w_2\cchi{\frac{n}{\alpha}}:=z$.  This is the desired number since
$z^\beta=w_1^\beta=z_1$ and $\eta^{-1}z^{-\alpha}=\eta^{-1}w_2^{-\alpha} = z_2$ and first equation of (\ref{zequation2}) becomes the first equation of (\ref{zequation3}).

Part (b) is a straightforward calculation.

Part (c).
The condition for a root $z$ of the Laurent polynomial in (\ref{zequation3}) to be a multiple root is
$
\,\frac{d}{dz} \left[ \eta z^\alpha + \eta^{-1}z^{-\alpha} + z^\beta + z^{-\beta} \right] \,\not=\, 0\,
$
at the root $z$, or
\begin{equation}\label{multipleroot}
  \beta\left( z^\beta - z^{-\beta} \right) = -\alpha\left( \eta z^\alpha - \eta^{-1} z^{-\alpha} \right).
\end{equation}
Both sides of this equation lie on ellipses in $\CC$ with vertical major axis.
The left side lies on an ellipse with major radius $\beta\left( |z|^\beta + |z|^{-\beta} \right)$ and minor radius $\beta\left| |z|^\beta - |z|^{-\beta} \right|$, and the right side lies on an ellipse with major radius $\alpha\left( |z|^\alpha + |z|^{-\alpha} \right)$ and minor radius $\alpha\big| |z|^\alpha - |z|^{-\alpha} \big|$.  The assumptions $\beta>\alpha$ and $|z|\not=1$ imply
$\beta\left( |z|^\beta + |z|^{-\beta} \right)>\alpha\left( |z|^\alpha + |z|^{-\alpha} \right)$
and $\beta\big| |z|^\beta - |z|^{-\beta} \big|>\alpha\big| |z|^\alpha - |z|^{-\alpha} \big|$,
so that the two ellipses do not intersect.  Thus (\ref{multipleroot}) cannot hold when $\left| z \right|\not=1$, and such $z$ is therefore a simple~root of (\ref{zequation3}).
\end{proof}

The full set of solutions of the dispersion relation (\ref{zequation1}) for the tube is the union of the disjoint sets of solutions to the equations (\ref{zequation2}) for $0\leq\ell<\delta$.

\begin{theorem}\label{thm:zequation}
(a) If $\alpha$ and $\beta$ are positive integers with $\gcd(\alpha,\beta)=1$ and $\delta$ is a positive integer, then
the set of solutions $(z_1,z_2)\in(\CC^*)^2$ to the system (\ref{zequation1}) is the {\em disjoint} union
\begin{equation}
  \bigcup\limits_{\ell=0}^{\delta-1} {\mathcal Z}_\ell
\end{equation}
of the solutions sets of (\ref{zequation2}),
\begin{equation}\label{union}
  {\mathcal Z}_\ell \,=\,
  \left\{ (z^\beta, \eta^{-1}z^{-\alpha})\,:\,
    z^\beta + z^{-\beta} + \eta z^\alpha + \eta^{-1} z^{-\alpha} = 4\,c(\lambda,1),\; z\in\CC^*,\; \eta=\cchi{\frac{-\ell}{\beta\delta}}
  \right\}.
\end{equation}

(b) Using the set operation notation $f({\mathcal Z})=\left\{ \big(f(z_1),f(z_2)\big) \,:\, (z_1,z_2)\in{\mathcal Z} \right\}$ with $f=\cdot^{-1}$ or $f=\bar\cdot$,
one has
\begin{eqnarray}
  && {\mathcal Z}_0^{-1} = \overline{\mathcal Z}_0 = {\mathcal Z}_0\,, \\
  && {\mathcal Z}_\ell^{-1} = \overline{\mathcal Z}_\ell = {\mathcal Z}_{\delta-\ell} \,.
\end{eqnarray}
Thus, for each $\ell$, ${\mathcal Z}_\ell$ is invariant under reflection about the unit circle $(z_1,z_2)\mapsto(\bar z_1^{-1},\bar z_2^{-1})$, and ${\mathcal Z}_{\delta/2}$ (when $\delta$ is even) and ${\mathcal Z}_0$ are invariant under reflection about the real axis.
\end{theorem}

\begin{proof}
(a) Because the system (\ref{zequation1}) is equivalent to the existence of an integer $\ell:0\leq\ell<\delta$ such that (\ref{zequation2}) holds, Lemma~\ref{prop:zequation} establishes the union (\ref{union}).  To prove that it is disjoint, suppose that
\begin{equation}
  \renewcommand{\arraystretch}{1.1}
\left. 
  \begin{array}{r}
    z_1=z^{\beta} = w^{\beta} \\
    z_2 = \eta^{-1} z^{-\alpha} = \nu^{-1} w^{-\alpha}
  \end{array}
\right\}
\quad\text{with $\eta=\cchi{\frac{-\ell}{\beta\delta}}$ and $\nu=\cchi{\frac{-k}{\beta\delta}}$}
\end{equation}
and $0\leq\ell<\delta$ and $0\leq k<\delta$.
One has $\eta\nu^{-1}=(wz^{-1})^\alpha$ and $(wz^{-1})^\beta=1$, so that
$(\eta\nu^{-1})^\beta = (wz^{-1})^{\alpha\beta} = 1^\alpha = 1$, and therefore $\cchi{\frac{k-\ell}{\delta}}=1$.  Since $|k-\ell|<\delta$, one has $\left| \frac{k-\ell}{\delta} \right|<1$, which together with $\cchi{\frac{k-\ell}{\delta}}=1$ yields $\frac{k-\ell}{\delta}=0$ and hence $\eta=\nu$.

Part (b) of the theorem is a result of part (b) of Lemma~\ref{prop:zequation}.
\end{proof}

\begin{proposition}\label{prop:fluxes}
Let $(z_1,z_2)$ and $(w_1,w_2)$  be solutions of the dispersion system (\ref{zequation1}) with $\lambda\not\in\sigma_D$, and let $u$ and $v$ from $V_\lambda$ be corresponding eigenfunctions.

1.  If $|z_1|=|z_2|=1$ and $(w_1,w_2) = (\bar z_1,\bar z_2)$ and these are simple solutions to (\ref{zequation3}), then $[u,u]$ and $[v,v]$ are nonzero real numbers of opposite sign.

2.  If $(w_1,w_2) = (\bar z_1^{-1},\bar z_2^{-1})$, then $[u,v]\not=0$.
\end{proposition}

\begin{proof}
The flux $[\underline u,\,\underline v]_\lambda$, which is computed as the sum of the fluxes along edges crossing the loop $L$ can also be computed as the sum of the fluxes along edges 
emanating from the cycle $C$ illustrated in Fig.~\ref{fig:cycle}, and directed into the right side of $C$.  This flux-conservation law is due to the Neumann vertex conditions satisfied by $u$ and $v$ (see~(\ref{domA1})).
Let $E_C$ denote all the edges emanating rightward or downward into the region to the right of $C$.

\begin{figure}  
\centerline{\scalebox{0.27}{\includegraphics{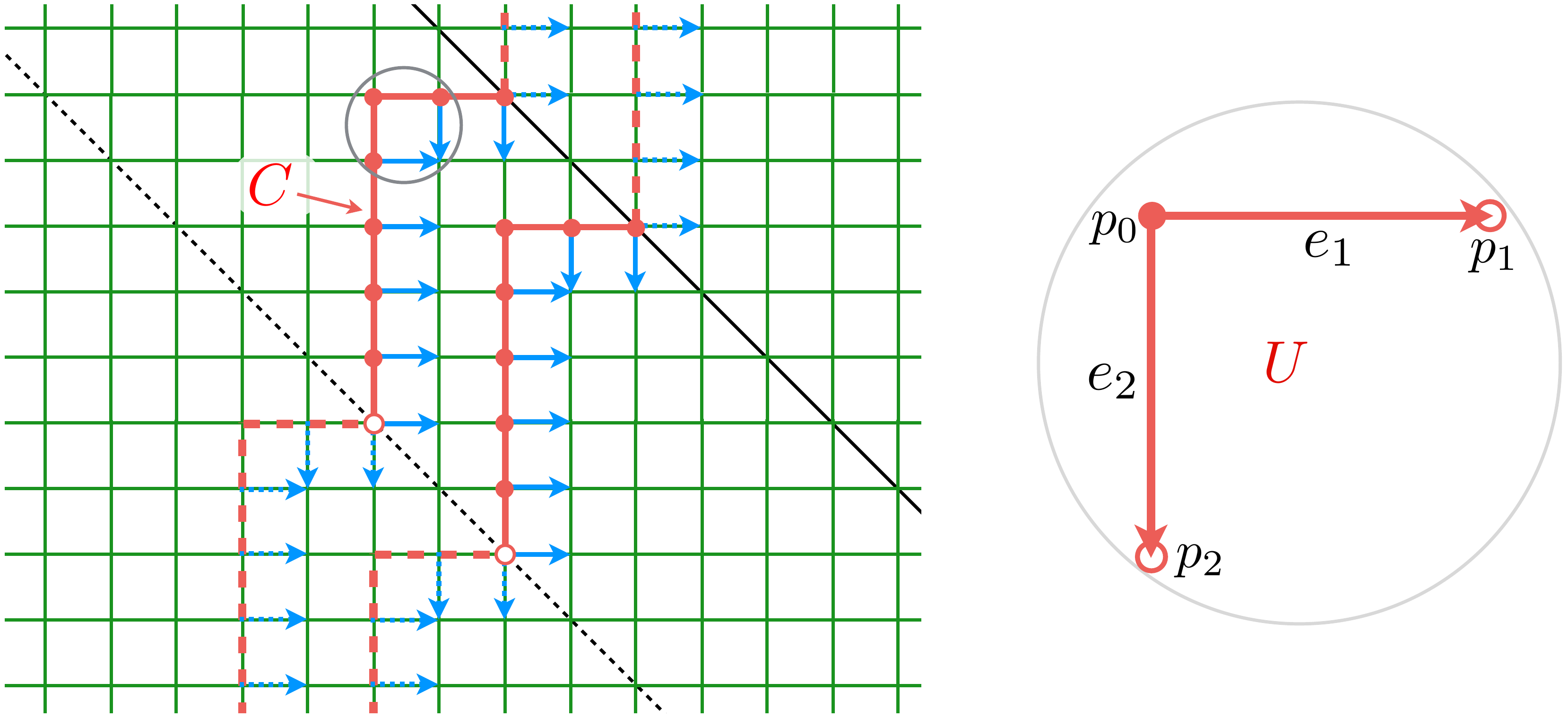}}}
\caption{\small The cycle $C$ around the tube $T$ used in the proof of Proposition~\ref{prop:fluxes} and the fundamental domain $U$ consisting of the vertex $p_0$ and the edges $e_1$ and $e_2$.}
\label{fig:cycle}
\end{figure}

Let $u$ and $v$ be normalized so that $u(p_0)=v(p_0)=1$.  Then $u(p_1)=z_1$, $u(p_2)=z_2^{-1}$, $v(p_1)=w_1$, and $v(p_2)=w_2^{-1}$.  From these values and the expression (\ref{flux2}) for the flux along an edge, one obtains
\begin{eqnarray}
  s(\lambda,1) (\bar u v' - \bar u' v)|_{e_1} &=& (w_1-\bar z_1), \\
  s(\lambda,1) (\bar u v' - \bar u' v)|_{e_2} &=& (w_2^{-1}-\bar z_2^{-1}).
\end{eqnarray}
By using the Floquet relations $\mfh u = z_1 u$ and $\mfv u = z_2 u$, and likewise for $v$, one obtains
\begin{equation}
\begin{split}
  2i\,s(\lambda,1)\,[\underline u,\underline v]
  &= s(\lambda,1) \sum_{e\in E_C} (\bar u v' - \bar u' v)|_e \\
  & =  (w_2^{-1}-\bar z_2^{-1}) \sum_{j=1}^{\alpha\delta} (\bar z_1w_1)^j
     + (w_1-\bar z_1) \sum_{j=1}^{\beta\delta} (\bar z_2w_2)^{-j}
\end{split}
\end{equation}
If $(w_1,w_2) = (\bar z_1^{-1}, \bar z_2^{-1})$, one obtains $\bar z_1 w_1 = \bar z_2 w_2 = 1$, so that this relation simplifies to
\begin{equation}\label{hello}
\begin{split}
  2i\,s(\lambda,1) \overline{[\underline u,\underline v]}
         &\,=\, \alpha\delta (z_2^{-1} - z_2) + \beta\delta (z_1 - z_1^{-1}) \\
         &\,=\, \delta \left( \alpha \left( \eta z^\alpha - \eta^{-1}z^{-\alpha} \right)
               + \beta \left( z^\beta - z^{-\beta} \right) \right) \\
         &\,=\, \delta\, z \frac{d}{dz} \left[ \eta z^\alpha + \eta^{-1}z^{-\alpha} + z^\beta + z^{-\beta} \right].
\end{split}
\end{equation}
In case (2) of the Proposition, for which $|z|\not=1$, this quantity does not vanish because of part (c) of Proposition~\ref{prop:zequation}.  (In fact, case (2) also follows from Theorem~\ref{thm:modes} and part (c) of Proposition~\ref{prop:zequation}.)

For case (1) of the Proposition, put $|z|=1$ in (\ref{hello}), so that $(w_1,w_2)=(z_1,z_2)$ and $u=v$.  The assumption that this is a simple solution of (\ref{zequation3}) makes (\ref{hello}) nonzero.  Upon replacing $(w_1,w_2)$ and $(z_1,z_2)$ with $(\bar w_1,\bar w_2)=(w_1^{-1},w_2^{-1})$ and $(\bar z_1,\bar z_2)=(z_1^{-1},z_2^{-1})$, the expression (\ref{zequation3}) changes sign.
\end{proof}

\subsection{Spectrum of the tube}

The spectrum of the operator $A$ on the tube $T$ is the set $\sigma(T)=\sigma_\text{ac}\cup\sigma_D$, where
\begin{equation}\label{Tspectrum}
  \sigma_\text{ac} \,=\, \left\{ \lambda\in\RR\setminus\sigma_D : \exists \text{ solution $(z_1,z_2)$ of (\ref{zequation1}) with $|z_1|=|z_2|=1$} \right\}\,.
\end{equation}
This absolutely continuous part of the spectrum is composed of multiple bands determined by the real part of the dispersion relation for the tube, each band having even multiplicity.  These bands correspond to the pairs Floquet modes with unit-modulus Floquet multiplier in Theorem~\ref{thm:modes}.  The reader is referred to \cite{KuchmentPost2007} for an in-depth analysis of the spectrum of quantum graph tubes, specialized to the hexagonal case.  A result similar to Theorem~4.3 in that work holds here.  In this section, it will suffice to describe how the bands arise through the Fourier, or Floquet, transform.

\begin{figure}  
\centerline{\scalebox{0.52}{\includegraphics{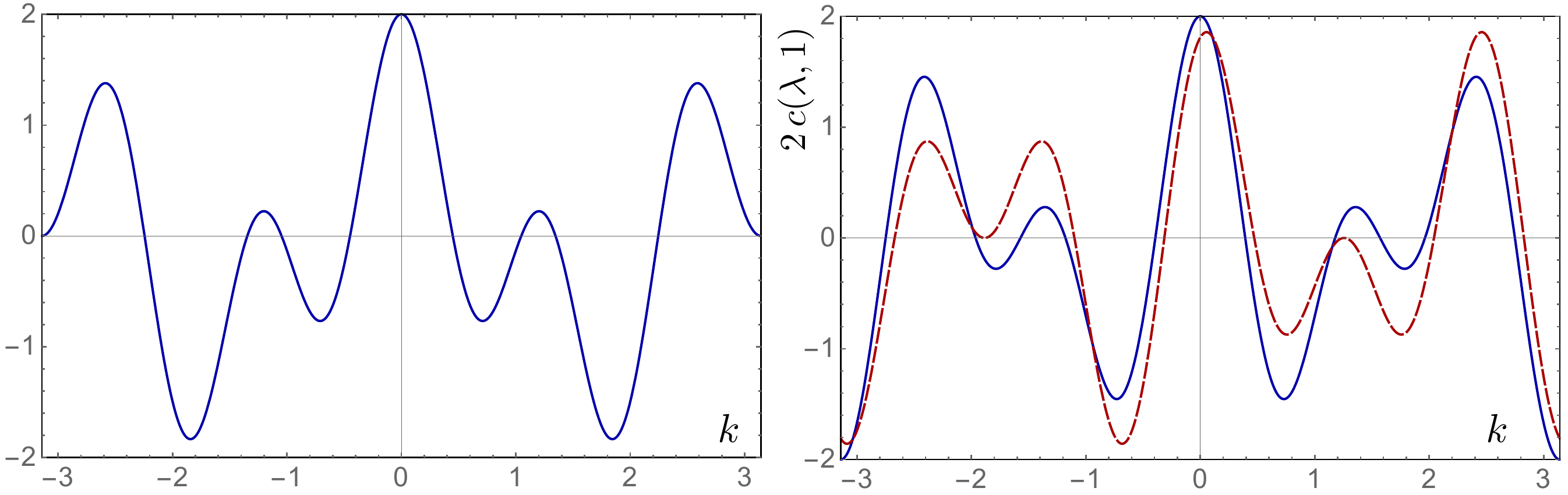}}}
\caption{\small The spectral bands of the operator $A=-\partial_{xx}$ on the tube $T$ are depicted for $\alpha=2$, $\beta=5$, $\delta=1$, on the left; and for $\alpha=3$, $\beta=5$, $\delta=2$ on the right.  The spectral bands are the $\lambda$-intervals that correspond to the monotonic segments of the graphs $\cos(\beta k) + \cos(\alpha k - 2\pi\ell/(\beta\delta)) = 2\,c(\lambda,1)$\, for $0\leq\ell<\delta$.  Here, $q(x)=0$ so that $c(\lambda,1)=\cos\sqrt{\lambda\,}$.
}
\label{fig:spectrum}
\end{figure}

A graphical depiction of the spectrum of the tube is obtained from the characterization (\ref{zequation3}) of the dispersion relation.  By putting $z=e^{ik}$, one obtains
\begin{multline}
  \sigma_\text{ac} \,=\,
        \big\{ \lambda\in\RR : \text{$\exists\;k\in(-\pi,\pi]$ and $\ell\in\{0,\dots,\delta\!-\!1\}$ s.th.}\; \\
                              \cos(\beta k) + \cos(\alpha k - 2\pi\ell/(\beta\delta)) = 2\,c(\lambda,1) \big\}.
\end{multline}
Fig.~\ref{fig:spectrum} depicts the relations $\cos(\beta k) + \cos(\alpha k - 2\pi\ell/(\beta\delta)) = 2\,c(\lambda,1)$.  Each monotonic segment of these graphs corresponds to a sequence of spectral bands given by $2\,c(\lambda,1)\in[a,b]$, where $[a,b]$ is the range of the segment.  There are $2\beta\delta$ sequences of bands.

The energies $\lambda$ marking the edges of spectral bands (extremal points on the graphs in Fig.~\ref{fig:spectrum}) form a set that has no accumulation points.
Except at these energies, there are $2\beta\delta$ geometrically simple Floquet multipliers, all corresponding to Floquet modes as described in Theorem~\ref{thm:modes}.
At a band edge energy $\lambda$, two Floquet multipliers merge and the operator $P_\lambda$ has a nontrivial Jordan block.

\begin{figure}  
\centerline{\scalebox{0.4}{\includegraphics{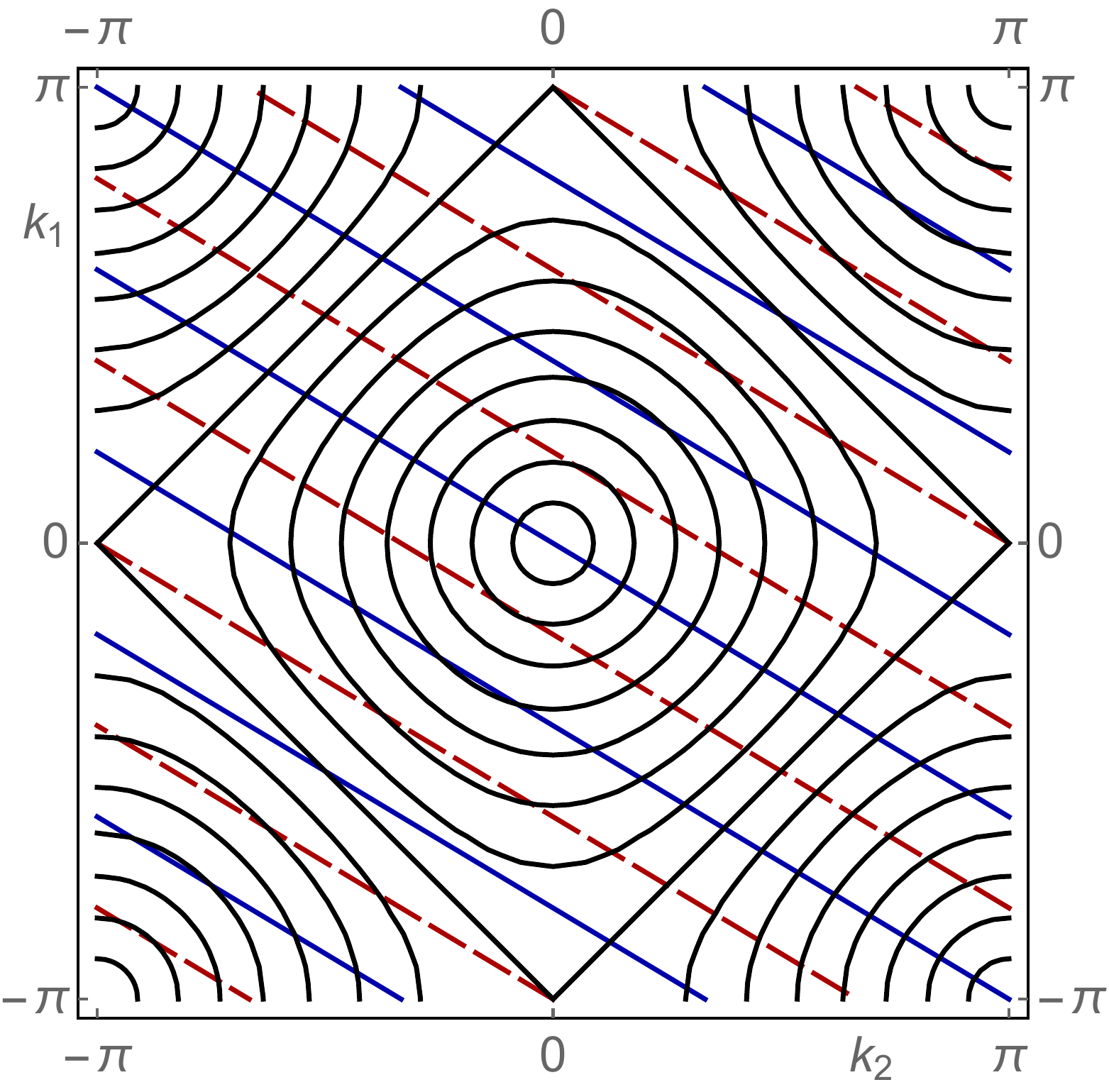}}}
\caption{\small
A depiction of the spectral bands of the operator $A=-\partial_{xx}$ on the tube $T$ for $\alpha=3$, $\beta=5$, $\delta=2$.  Shown are the level sets of the dispersion relation $\cos k_1 + \cos k_2 = 2\,\cos\sqrt{\lambda\,}$ for different values of $\lambda\in\RR$; and the linear restriction $\delta\alpha k_1 + \delta\beta k_2 \equiv 0$ (mod $2\pi$) for $\alpha=3$, $\beta=5$, and $\delta=2$.  The linear restriction consists of $\delta=2$ lines on the torus $(\RR/2\pi\ZZ)^2$ that traverse $\lambda$-intervals that form the spectral bands of the quantum tube $T$; one is in solid blue and the other in dashed red.  The graphs of $2\cos\sqrt{\lambda\,}$ along these lines are shown in Fig.~\ref{fig:spectrum} (right).
}
\label{fig:contours}
\end{figure}

Alternatively, as in \cite{KuchmentPost2007}, one can consider the dispersion relation (\ref{zequation2}) with $z_1=e^{ik_1}$ and $z_2=e^{ik_2}$ as a function of $k_1$ and $k_2$, namely, $\cos k_1 + \cos k_2 = 2\,c(\lambda,1)$, and restrict it to the relation $\delta\alpha k_1 + \delta\beta k_2 \equiv 0$ (mod $2\pi$), as shown in Fig.~\ref{fig:contours}.  This restriction amounts to $\delta$ lines on the torus $(\RR/2\pi\ZZ)^2$, each parameterized by $k$ and corresponding to one of the graphs in Fig.~\ref{fig:spectrum}.

\smallskip
The characterization (\ref{Tspectrum}) of the spectrum of $A$ on $T$ is understood through the Fourier transform with respect to the group $G$ of symmetries of $T$.  The group of complex characters for the abelian group $G$ is isomorphic to the subgroup of $(\CC^*)^2$ given by
\begin{equation}
  G^* \,=\, \left\{ (z_1,z_2) \in (\CC^*)^2 : z_1^{\alpha\delta} z_2^{\beta\delta} = 1 \right\},
\end{equation}
whereby $(z_1,z_2)\in G^*$ corresponds to the character $\mfh^m\mfv^n \mapsto z_1^m z_2^n$.
The Floquet transform of $u(x)$ is a variation of the Fourier transform.  It is a formal series in the variable $(z_1,z_2)\in G^*$ whose coefficients are shifts of $u$,
\begin{equation}\label{ft}
  \hat u(x,z_1,z_2) \,=\, \sum_{\mfh^m\mfv^n \in G} u(\mfh^m\mfv^n x) z_1^{-m}z_2^{-n}\,.
\end{equation}
The function $u$ is recovered through the inverse Floquet transform obtained by integrating the evaluation of $\hat u$ over the unitary representations $G^*\cap\TT^2$, where $\TT^2$ is the two-dimensional torus $\left\{ z\in\CC : |z|=1 \right\}^2$\,:
\begin{equation}\label{ftinv}
  u(\mfh^m\mfv^n\,x) \,=\, \frac{1}{(2\pi)^2} \int_{G^*\cap\TT^2} \hat u(x,e^{ik_1},e^{ik_2}) e^{i(k_1m+k_2n)} d\mu(k_1,k_2)\,,
\end{equation}
in which $d\mu$ is the Lebesgue measure on $G^*\cap\TT^2$ inherited from the metric of $\TT^2$.  In this integral, the group $G^*\cap\TT^2$ is viewed as
\begin{equation}
  \left\{ (k_1,k_2) \in \RR^2/(2\pi\ZZ)^2 : (\alpha k_1 + \beta k_1)\delta \equiv 0\; \text{(mod $2\pi$)} \right\}.
\end{equation}

One can see from (\ref{ft}) that for a given $(z_1,z_2)\in G^*$, $\hat u(\cdot,z_1,z_2)$ is a simultaneous eigenfunction of $\mfh$ and $\mfv$ (it is pseudo-periodic),
\begin{equation}\label{pseudo-periodic}
  \hat u(\mfh x,z_1,z_2) \,=\, z_1\,\hat u(x,z_1,z_2)\,,
  \qquad
  \hat u(\mfv x,z_1,z_2) \,=\, z_2\,\hat u(x,z_1,z_2)\,.
\end{equation}
In fact, the Floquet theory (the theory in \cite{ReedSimon1980d} extends to quantum graphs) guarantees that, if $u\in\mathrm{dom}(A)$ as stipulated in (\ref{domA2}), then $\hat u(\cdot,e^{ik_1},e^{ik_2})$ for real $k_1$ and $k_2$, as a true sum (not just formal), lies in $\Hloc(T)$ and possesses, in addition, the pseudo-periodicity property (\ref{pseudo-periodic}).
Thus, $\hat u(\cdot,e^{ik_1},e^{ik_2})$ is in the function space
\begin{equation}
  \Hloc(k_1,k_2) \,=\,
  \Bigg\{
     f \in \bigoplus_{e\in\mathcal{E}(T)} H^2(e) \,:
     \sum_{e\in\mathcal{E}_v(T)} f_e'(v) =0 \;\;\forall\,v\in\mathcal{V}(T),\;
     \mfh f = e^{ik_1}f,\;
     \mfv f = e^{ik_2}f
  \Bigg\}.
\end{equation}

Because functions in $\Hloc(k_1,k_2)$ are determined by their values on the fundamental domain $U$ for $G$ shown in Fig.~\ref{fig:cycle}, $\Hloc(k_1,k_2)$ can be identified with the space $H^2_{k_1,k_2}(U)$ of $H^2$ functions on $W$ with conditions at the boundaries of the two edges set by the pseudo-periodicity condition (\ref{pseudo-periodic}), and the function $\hat u(\cdot,e^{ik_1},e^{ik_2})$ for real $k_1$ and $k_2$ might as well be taken to be in $H^2_{k_1,k_2}(U)$.   If $f|_{e_1}=f_1$ and $f|_{e_2}=f_2$, this space is stipulated precisely by
\begin{multline}
  H^2_{k_1,k_2}(U) =
    \Big\{
     f \in H^2(e_1) \oplus H^2(e_2) \,:
     f_1(v_0) = f_2(v_0) = e^{-ik_1}f_1(v_1) = e^{-ik_2}f_2(v_2)\,, \\
     f_1'(v_0) + f_2'(v_0) - e^{-ik_1}f_1'(v_1) - e^{-ik_2}f_2'(v_2) = 0
  \Big\}.
\end{multline}
The Floquet transform is a Hilbert-space isomorphism from $L^2$ of the tube to $L^2$ of the torus with values in $L^2(U)$,
\begin{equation}
  \hat{}\; : L^2(T,\CC) \to L^2(\TT^2\cap G^*, L^2(U,\CC))\,.
\end{equation}

Under the Floquet transform, the operator $A$ retains its form as a differential operator,
\begin{equation}
  (Au)\hspace{1pt}\hat{}\,(x,e^{ik_1},e^{ik_2})
     = \big(\!-\partial_{xx}+q(x)\big)\, \hat u(x,e^{ik_1},e^{ik_2}),
\end{equation}
except that it is acting in the $(k_1,k_2)$-dependent domain $H^2_{k_1,k_2}(U)$.
Therefore $A-\lambda$ is invertible whenever $-\partial_xx + q(x) - \lambda$ is invertible on each of the domains $H^2_{k_1,k_2}(U)$ with $(e^{ik_1},e^{ik_2})\in G^*$.
The operator $-\partial_{xx}+q(x)$ on the domain $H^2_{k_1,k_2}(U)$ is self-adjoint in $L^2(U)$, and its spectrum is discrete.  It is equal to the set $\big\{ \lambda\in\RR : z_1 + z_1^{-1} + z_2 + z_2^{-1} = 4\,c(\lambda,1) \big\}$.  The union of these sets over $(z_1,z_2)\in G^*\cap \TT^2$ is the spectrum of the quantum-tube operator~$A$, as given in (\ref{Tspectrum}).

\section{The semi-infinite tube}\label{sec:HalfTube} 

Let $\tilde T_+$ denote the half-tube as a metric graph obtained by retaining only the non-negative $\mfh$-translates of~$W$,
\begin{equation}
  \tilde T_+ \,=\, \bigcup\limits_{m=0}^\infty \mfh^m W\,.
\end{equation}
Then define $T_+$ by augmenting $\tilde T_+$ by attaching a finite metric graph $\Gamma_\text{aux}$ with vertex set $\mathcal{V}_\text{aux}$, containing $\gamma$ vertices.  The graphs are attached by introducing finitely many edges that connect vertices of $\Gamma_\text{aux}$ with the $\delta\beta$ boundary vertices of $\tilde T_+$, which are the $\delta\beta$ vertices of $W$, denoted by $v_n : 0\leq n<\beta\delta$.    Denote the vertices in $\mathcal{V}_\text{aux}$ that are connected to the boundary vertex $v_n$ of $\tilde T_+$ by $\{ {v_{ni} : 1\leq i\leq \gamma_n} \}$\,.  Let the edge connecting $v_{ni}$ to $v_n$ be have length $L_{ni}$, and let the symmetric potential on this edge be denoted by $q_{ni}(x)$.  Let each vertex of $\Gamma_\text{aux}$ and the vertices $v_n : 0\leq n<\beta\delta$ be equipped with a Robin condition
\begin{equation}\label{Robin}
  a f(v) \,+\, b \sum_{e\text{ inc }v} f_e'(v) \:=\: 0\,,
\end{equation}
in which $a$ and $b$ are real numbers and the derivatives are taken to be directed away from the vertex $v$.
Extend $A$ to $T_+$ by letting it act by $-\partial_{xx} + q_e(x)$ on each edge and requiring functions in the domain of $A$ to be in $H^2$ of each new edge and satisfy the specified Robin vertex conditions.  These vertex conditions render $A$ self-adjoint in $L^2(T_+)$ \cite{BerkolaikoKuchment2013}.  Denote by $\sigma_D(T_+)$ the Dirichlet spectrum of $T_+$, which consists of the union of the Dirichlet spectra of all of the edges of $T_+$.

The spectrum $\sigma(T_+)$ of $A$ on $T_+$ contains $\sigma(T)$, and, as such, each $\lambda\in\sigma_D$ continues to have infinite multiplicity and the continuous part of $\sigma(T_+)$ is still $\sigma_\text{ac}$, but the multiplicity for $T_+$ is half of the multiplicity for the full tube $T$.  The extended states corresponding to the continuous spectrum are states that are reflected, or scattered, by the boundary of the half-tube.
The half-tube may also have additional point spectrum $\sigma_\text{pp}$ of finite multiplicity corresponding to bound states that decay exponentially but have infinite support.  Thus, the spectrum of the half-tube is
\begin{equation}
  \sigma = \sigma_\text{ac} \cup \sigma_\text{pp} \cup \sigma_D\,.
\end{equation}

The spectrum of a quantum tube perturbed by a localized defect is of this same form, although the multiplicity of $\sigma_\text{ac}$ remains the same as that for the full unperturbed tube.  A full proof of the spectrum of the perturbed tube follows the lines of argument in \cite{Goldstein1969} for the Laplacian in a locally perturbed infinite waveguide.  The proof for the semi-infinite tube is similar; the analysis is carried out in detail for zigzag hexagonal quantum tubes in \cite{IantchenkoKorotyaev2010}.  Here, it will suffice to describe the reflection problem and prove that point spectrum, including embedded eigenvalues, is possible.

\subsection{Reflection of modes in the half-infinite tube}

Let $u\in\Hloc(T_+)$ satisfy $(A-\lambda)u=0$.  The restriction of $u$ to $\tilde T_+$ is a combination of the Floquet modes of the tube,
\begin{equation}\label{modesum}
  u|_{\tilde T_+} \,=\, \sum_{j=1}^{\beta\delta} \left( c_{j+}u_{(j+)} + c_{j-}u_{(j-)} \right).
\end{equation}
The curve wrapping around $\tilde T_+$, formed by the line $L$ introduced in the definition of the flux $[\cdot,\,\cdot]$, bounds a finite part of the graph $T_+$ that contains the attachment $\Gamma_\text{aux}$.  
The self-adjoint vertex conditions yield a conservation law, namely that $[\underline u,\,\underline v]=0$ for any pair of $\lambda$-eigenfunctions of $A$ on $T_+$.
The flux relations (\ref{modefluxes}) between pairs of modes given in Proposition~\ref{thm:modes} imply
\begin{equation}\label{conservation}
  0 \,=\, [\underline u(0),\underline u(0)] \,=\,
  \sum_{|z_{j}|=1} \left( |c_{j+}|^2 - |c_{j-}|^2 \right) + 2 \sum_{|z_{j}|<1} \Re\,\bar c_{j+}c_{j-}\,.
\end{equation}

As long as $\lambda\not=\sigma_D(T_+)$, the values of $u$ on the vertices determine $u$ and therefore also the derivatives of $u$ at the vertices.  The vertex conditions can therefore be written in terms of the values of $u$ on the vertices.  The mode sum (\ref{modesum}) satisfies the Neumann vertex conditions at internal vertices of $T_+$, that is, excluding $\left\{ v_k : 0\leq k<\beta\delta \right\}$.
Imposing the vertex conditions of the form (\ref{Robin}) at the $\gamma+\beta\delta$ vertices of the set $\mathcal{V}_\text{aux} \cup \left\{ v_k : 0\leq k<\beta\delta \right\}$ results in $\gamma+\beta\delta$ homogeneous linear equations in the $\gamma+2\beta\delta$ quantities $\left\{ u(v) : v\in\mathcal{V}_\text{aux} \right\}$ and $\left\{ c_{j+},\,c_{j-} \right\}_{j=1}^{\beta\delta}$.
This means that these quantities lie in the nullspace of a linear map
\begin{equation}\label{F}
  F \,:\,  \CC^\gamma \times \CC^{\beta\delta} \times \CC^{\beta\delta} \to \CC^{\gamma+\beta\delta},
\end{equation}
in which the first coordinate is $\left\{ u(v) : v\in\mathcal{V}_\text{aux} \right\}$, the second is
$\left\{ c_{j+} \right\}_{j=1}^{\beta\delta}$ (outgoing modes), and the third is $\left\{ c_{j-} \right\}_{j=1}^{\beta\delta}$ (incoming modes).
The fields $u\in\Hloc(T_+)$ associated with the nullspace of $F$ are all the solutions to $(A-\lambda)u=0$ on the half-infinite tube. 

If the restriction of $F$ to the first two coordinates,
\begin{equation}\label{Frestricted}
  F \,:\,  \CC^\gamma \times \CC^{\beta\delta} \times \left\{ 0 \right\} \to \CC^{\gamma+\beta\delta},
\end{equation}
is invertible, then there exists a linear function
\begin{equation}
  S : \CC^{\beta\delta} \to \CC^{\beta\delta}
     :: \left\{ c_{j-} \right\}_{j=1}^{\beta\delta} \mapsto \left\{ c_{j+} \right\}_{j=1}^{\beta\delta}
\end{equation}
such that
\begin{equation}\label{scattering}
  F\left( \left\{ u(v) : v\in\mathcal{V}_\text{aux} \right\},\; S \left\{ c_{j-} \right\}_{j=1}^{\beta\delta},\,\left\{ c_{j-} \right\}_{j=1}^{\beta\delta}\, \right) \,=\, 0 \,.
\end{equation}
$S$ is the {\em scattering operator} associated with reflection of leftward modes from the boundary of $T_+$ back into rightward modes.
Solutions to (\ref{scattering}) with $c_{j-}\not=0$ for some $j$ but $c_{j-}=0$ for all $j$ such that $|z_{j-}|\not=1$ are the modified extended states associated with the continuous spectrum.

The relations defining $F$ look complicated when written explicitly, so set $\delta=1$ for simplicity.
Let the Robin conditions at the vertices $\left\{ v_n : 0\leq n<\beta \right\}$ be given as
\begin{equation}\label{robinvn}
  a_n u(v_n) + b_n \Big( \sum_{e\text{ inc }v} u_e'(v_n) \Big) \,=\, 0.
\end{equation}
By normalizing the modes $u_{(j\pm)}$ so that $u_{(j\pm)}(v_0)=1$, one has $u_{(j\pm)}(v_n) = z_{(j\pm)}^{r(n)}$, where $r(n)=s(n)\beta-n\alpha$ and $s(n)$ is the number of horizontal shifts required to pass from $v_0$ to~$v_n$ and $z_{(j\pm)}$ is the $z$-value that corresponds to $u_{(j\pm)}$ as described in Proposition~\ref{prop:zequation}.

For the edge $e=(v,w)$ connecting $v$ and $w$,
let $s(\lambda,v,w)$ denote $u(w)$ where $u(x)$ is the solution to $(-\partial_{xx}+q_e(x)-\lambda)u=0$ such that $u(v)=0$ and $u'(v)=1$, and let $c(\lambda,v,w)$ denote $u(w)$ for the solution with $u(v)=1$ and $u'(v)=0$.
  Set $\chi_n=0$ if $v_n$ is incident to exactly two edges of $W$, and set $\chi_n=1$ if $v_n$ is incident to three edges of $W$.
The conditions (\ref{robinvn}) for $0\leq n<\beta$, written as a homogeneous linear equation in the quantities
$\left\{ u(v) : v\in\mathcal{V_\text{aux}} \right\}$ and $\left\{ c_{j+},\,c_{j-} : 1\leq j\leq\beta \right\}$, is
\begin{multline}\label{F1}
  0 \:=\: a_n \sum_{j=1}^\beta \left( c_{j+}z_{(j+)}^{r(n)} + c_{j-}z_{(j-)}^{r(n)} \right) + \\
     \;+\; b_n\frac{1}{s(\lambda,1)}
            \sum_{j=1}^\beta \left[ c_{j+}z_{(j+)}^{r(n)} \left( z_{(j+)}^\beta+z_{(j+)}^\alpha-2c(\lambda,1) + (z_{(j+)}^{-\alpha}-c(\lambda,1))\chi_n \right) \,+\, \right. \\
      \hspace{8em} \left.     \,+\; c_{j-}z_{(j-)}^{r(n)} \left( z_{(j-)}^\beta+z_{(j-)}^\alpha-2c(\lambda,1) + (z_{(j-)}^{-\alpha}-c(\lambda,1))\chi_n \right) \right] \;+\; \\
      +\;  b_n\hspace{-1em} \sum_{\mbox{\parbox{6em}{\scriptsize\centering $v\in\mathcal{V}_\text{aux}$\\ $(v,v_n)\in\mathcal{E}(T_+)$}}} \hspace{-1em}
      \frac{1}{s(\lambda,v,v_n)}
      \left[ u(v) - c(\lambda,v,v_n)) \sum_{j=1}^\beta \left( c_{j+}z_{(j+)}^{r(n)} + c_{j-}z_{(j-)}^{r(n)} \right) \right].
\end{multline}
The $\gamma$ Robin conditions at the vertices $v\in\mathcal{V}_\text{aux}$
\begin{equation}\label{robinv}
  a_v u(v) + b_v \Big( \sum_{e\text{ inc }v} u'_e(v) \Big) \;=\; 0\,
\end{equation}
are expanded as
\begin{multline}\label{F2}
  a_v u(v) \,+\, b_v\hspace{-2pt} \sum_{\mbox{\parbox{4.8em}{\scriptsize\centering $w\in\mathcal{V}_\text{aux}$\\ $(v,w)\in\mathcal{E}(T_+)$}}} \hspace{-2pt} \frac{1}{s(\lambda,v,w)} \big( u(w) - c(\lambda,v,w) u(v) \big) \;+ \\
  +\;  b_v \sum_{(v_n,v)\in\mathcal{E}(T_+)} \frac{1}{s(\lambda,v_n,v)}
  \left( \sum_{j-1}^\beta \left( c_{j+} z_{(j+)}^{r(n)} + c_{j-} z_{(j-)}^{r(n)}  \right) - c(\lambda,v,v_n) u(v) \right)\,.
\end{multline}

\subsection{Bound states and point spectrum}

When the map (\ref{Frestricted}) is not invertible, there is a nonzero function $u\in\Hloc(T_+)$ satisfying $(A-\lambda)u=0$ such that $c_{j-}=0$ for all $j$ in the expansion (\ref{modesum}), that is, all of the leftward (incoming) modes of $u$ vanish.  The conservation law (\ref{conservation}) then implies that all of the rightward (outgoing) modes of $u$ with $|z_{(j+)}|=1$ also vanish,
\begin{equation}
  c_{j+} \,=\, 0
  \qquad
  \text{whenever} \; |z_{(j+)}|=1. 
\end{equation}
As a consequence, $u$ is exponentially decaying along the tube (recall $|z_{(j+)}|\leq1$ for all $j$) and is therefore an $L^2$ eigenfunction of $A$, or a {\em bound state} supported by the modified left end of the truncated tube.

The conditions for a bound state are (\ref{F1},\ref{F2}) with all $c_{j-}$ set to zero.
If $\lambda\not\in\sigma_c(T_+)$, then $|z_{(j+)}|<1$ for all $j$ and the resulting system for 
$\left\{ u(v) : v\in\mathcal{V_\text{aux}} \right\}$ and $\left\{ c_{j+} : 1\leq j\leq\beta \right\}$ is 
$(\gamma+\beta\delta)\times(\gamma+\beta\delta)$.  One expects it to be singular for a discrete set of $\lambda\not\in\sigma_c(T_+)$.

If $\lambda\in\sigma_c(T_+)$, then $|z_{(j+)}|=1$ for some values of $j$.  The coefficients $c_{j+}$ automatically vanish for these propagating modes; the system of equations for a bound state is therefore overdetermined, and the eigenvalue for this bound state would be embedded in the continuous spectrum.  One expects the system to be solvable for some frequencies for special choices of the terminating graph $\Gamma_\text{aux}$ and the coefficients in the Robin vertex conditions.

In the case that $\Gamma_\text{aux}$ is empty and $\delta=1$, the matrix of coefficients for the linear equations for the $c_{j+}$ is
\begin{equation}
  \left( z_{(j+)}^{r(n)} \left[ a_n + b_n\frac{1}{s(\lambda,1)} \left( z_{(j+)}^\beta + z_{(j+)}^\alpha - 2c(\lambda,1) + \big(z_{(j+)}^{-\alpha}-c(\lambda,1)\big)\chi_n \right) \right] \right)_{nj}\,.
\end{equation}
When $z_{(j+)}$ is real for some $j$, one can choose the coefficients $a_n$ and $b_n$ such that the $nj$-entries vanish for all $n$, and thus the mode $u_{(j+)}$ will satisfy the Robin conditions on the boundary vertices of $T_+$ and is thus a bound state.

When the potential on the edges is zero, one can make a clean statement about embedded and non-embedded bound states.  This proposition can be modified for nonzero potential in a straightforward way.
Note that any Floquet mode that participates in a bound state has $|z_1|<1$ and $|z_2|>1$, which corresponds to decay along the infinite direction of the half-tube.

\begin{proposition}[Bound states]\label{prop:boundstates}
  Let $T_+$ be the quantum graph consisting of the truncated tube $\tilde T_+=\cup_{m=0}^\infty\mfh^mW$  with the operator $A$ acting by $-\partial_{xx}$ on the edges and having domain subject to the Robin conditions (\ref{robinvn}) at the boundary vertices $v_n$ ($0\leq n<\beta\delta$) and the Neumann condition at all internal vertices.

In the following situations, the real numbers $a_n$ and $b_n$ for $0\leq n<\beta\delta$ can be chosen such that there exists a solution $u\in L^2(T_+)$ of $(A-\lambda)u=0$ that consists of a single exponentially decaying Floquet mode.  This solution is a bound state corresponding to the eigenvalue $\lambda$ of~$A$.  In each case, it is assumed that $\lambda\not=(k\pi)^2$ for all $k\in\ZZ$ and $\lambda$ not be a band edge.  (As always, $\alpha<\beta$ and $\gcd(\alpha,\beta)=1$.)

\renewcommand{\theenumi}{\alph{enumi}}
\begin{enumerate}

\item $\lambda<0$; $0<z_1<1$, and $1<z_2$;

\item $\beta$ is even; $\lambda<0$; $0<z_1<1$, $z_2<-1$;

\item $\beta$ is even; $\lambda>0$ with $\sqrt{\lambda}\in(-\pi/2,\,\pi/2)+2\pi\ZZ$; $0<z_1<1$, $z_2<-1$;

\item $\beta$ is odd; $\lambda>0$ with $\sqrt{\lambda}\in(\pi/2,\,3\pi/2)+2\pi\ZZ$; $-1<z_1<0$, $1<z_2$;

\end{enumerate}

\noindent
In cases (c) and (d), the eigenvalue $\lambda$ is embedded in the continuous spectrum, and in cases (a) and (b), it is not embedded.
\end{proposition}

\begin{proof}
When $q(x)=0$ for all $x\in[0,1]$, one has \,$s(\lambda,1)=\lambda^{-1/2}\sin\sqrt{\lambda}$\, and \,$c(\lambda,1)=\cos\sqrt{\lambda}$\, and $\sigma_D(T_+)=\left\{ (k\pi)^2 : k\in\ZZ \right\}$.
The first equation of (\ref{zequation3}) with $\eta=0$ becomes
\begin{equation}\label{reqn}
  f(z) := z^\beta + z^{-\beta} + z^\alpha + z^{-\alpha} \,=\, 4\cos\sqrt\lambda\,.
\end{equation}
The conditions for the Floquet mode corresponding to $z$ to satisfy the Robin conditions on the boundary vertices of $T_+$ are
\begin{equation}\label{robinz}
  a_n + b_n\frac{\sqrt\lambda}{\sin\sqrt\lambda} \left( z^\beta + z^\alpha - 2\cos\sqrt\lambda + \big(z^{-\alpha}-\cos\sqrt\lambda\big)\chi_n \right) \,=\, 0\,,
  \quad
  0\leq n<\beta\delta.
\end{equation}

The range of $f|_{(0,1)}$ is $f[(0,1)]=(4,\infty)$, and therefore for each $\lambda<0$, there is a solution of (\ref{reqn}) with $0<z<1$.
Let $\beta$ be even and $\alpha$ odd.  Then $f[(-1,0)]=(0,\infty)$, so for each $\lambda<0$, there is a solution of (\ref{reqn}) with $-1<z<0$.
And if $\lambda>0$ and $\sqrt\lambda\in(-\pi/2,\,\pi/2)+2\pi\ell$, then $4\cos\sqrt\lambda>0$ and there is a solution $z$ of (\ref{reqn}) with $-1<z<0$.  Now let $\beta$ be odd and $\alpha$ even.  Then $f[(-1,0)]=-(0,\infty)$, so if $\mu\in(\pi/2,\,3\pi/2)+2\pi\ell$, then $4\cos\sqrt\lambda<0$ and again there is a solution $z$ of (\ref{reqn}) with $-1<z<0$.

In each of these cases, since $z$ is real, (\ref{robinz}) is satisfied for appropriate choices of real numbers $a_n$ and~$b_n$.  The properties of $z_1$ and $z_2$ come from $z_1=z^\beta$ and $z_2=z^{-\alpha}$.
\end{proof}

\bigskip
\bigskip

\noindent
{\bfseries\large Acknowledgment.}  
This work was supported by NSF Research Grant DMS-1411393 (SPS) and NSF VIGRE grant 0739382 (JT).

\end{document}